# Quantum Interactions in Topological R166 Kagome Magnet

Xitong Xu[1]†, Jia-Xin Yin[2], Zhe Qu[1,3]†, Shuang Jia[4,5]†

[1]Anhui Key Laboratory of Condensed Matter Physics at Extreme Conditions, High Magnetic Field Laboratory, Hefei Institutes of Physical Science, Chinese Academy of Sciences, Hefei, Anhui 230031, China

[2]Department of Physics, Southern University of Science and Technology, Shenzhen 518055, China

[3]CAS Key Laboratory of Photovoltaic and Energy Conversion Materials, Hefei Institutes of Physical Science, Chinese Academy of Sciences, Hefei, Anhui 230031, China

[4]International Center for Quantum Materials, School of Physics, Peking University, Beijing 100871, China

[5]Interdisciplinary Institute of Light-Element Quantum Materials and Research Center for Light-Element Advanced Materials, Peking University, Beijing 100871, China

†Corresponding authors, E-mail: xuxitong@hmfl.ac.cn; zhequ@hmfl.ac.cn; gwljiashuang@pku.edu.cn

**Kagome magnet has been found to be a fertile ground for the search of exotic quantum states in condensed matter. Arising from the unusual geometry, the quantum interactions in the kagome lattice give rise to various quantum states, including the Chern-gapped Dirac fermion, Weyl fermion, flat band and van Hove singularity. Here we review recent advances in the study of the R166 kagome magnet ($RT_6E_6$, R = rare earths; T = transition metals; and E = Sn, Ge, etc.) whose crystal structure highlights the transition-metal-based kagome lattice and rare-earth sublattice. Compared with other kagome magnets, the R166 family owns the particularly strong interplays between the *d* electrons on the kagome site and the localized *f* electrons on the rare-earth site. In the form of spin-orbital coupling, exchange interaction and many-body effect, the quantum interactions play an essential role in the Berry curvature field in both the reciprocal and real spaces of R166 family. We discuss the spectroscopic and transport visualization of the topological electrons hosted in the Mn kagome layer of $RMn_6Sn_6$ and the various topological effects due to the quantum interactions, including the Chern-gap opening, the exchange-biased effect, the topological Hall effect and the emergent inductance. We hope this work serves as a guide for future explorations of quantum magnets.**

Since the discovery of the quantum Hall effect in the 1980s, topology has gradually been realized to be a new global parameter describing electronic states in condensed matter physics[1-4]. This conception is essentially associated with an intrinsic material property: the Berry curvature – a quantum mechanical quantity that is related to the phase of electron wavefunctions and corresponds to the gauge field defined on the parameter space. Certain collective electronic excitations in real materials, including the ones in the form of Dirac, Weyl and Majorana solutions of the corresponding quantum field theory[2-16], are characterized by an integer invariant integrated from the Berry curvature flux. Because the existence of the topological quasi-particles can lead to exotic electromagnetic responses in topological materials, they are promising for new technology in future energy-efficient electronics and quantum information science.

In the search for new topological materials, lattice models have always been a concise yet instructive tool. Certain lattice wave systems in translationally invariant settings, including the Haldane's honeycomb lattice, the Sutherland's dice lattice, the Lieb lattice, and the kagome lattice, etc, have spectral bands that are linearly crossed or independent of momentum in the tight binding approximation. These non-trivial topological structures arise from either inherent symmetries or fine-tuned hopping parameters[17-25]. Among these models, the kagome lattice has been one of the most experimentally studied systems as it has vast material candidates. The kagome lattice, consisting of corner-sharing triangles (Fig. 1**a**), has been studied in the context of quantum physics for over 70 years[22]. Tight binding calculation shows that without spin-orbital coupling (SOC), there exist relativistic Dirac dispersions at $K$ and $K'$ in the Brillouin zone promoting non-trivial topology (Fig. 1**c**). The destructive quantum interference of localized states results in a flat band over the Brillouin zone, which, together with the van Hove singularities at $M$ point at 3/12 and 5/12 fillings, allows the system to display remarkable strongly interacting phases including flat band ferromagnetism, Wigner crystal, fractional quantum Hall effect, electronic instability induced charge density wave (CDW) and unconventional superconductivity[25-39]. If the geometric frustration in the triangular network is considered, the kagome lattice of the spins may bear various competing quantum states with minute energy difference from valence bond solid to quantum spin liquid[40,41]. The kagome lattice system has therefore, stood out as a platform for studying the interplay between frustrated geometry, correlation and topology.

Various exotic quantum states have been experimentally demonstrated in bulk kagome materials in which the 2-dimensional (2D) kagome layers stack along the third dimension. Depending on the strength of interlayer interaction, the kagome magnets manifest distinct topological states, from a 2D massive Dirac fermion in $Fe_3Sn_2$ (Fig. 1**d**) to 3D Dirac and Weyl semimetals $Mn_3Sn$ and $Co_3Sn_2S_2$ [42, 44, 45]. The Berry curvature field introduces large anomalous Hall and Nernst effects in their electrical and thermoelectric transport[46–56]. The flat band with negative orbital magnetism has been identified by scanning tunnelling experiments in $Co_3Sn_2S_2$ (Fig. 1**e**) [43]. The van Hove singularities, which are closely related to the CDW and the unconventional superconductivity, were also observed in the kagome superconductor $AV_3Sb_5$, where A = K, Cs, Rb (Fig. 1**f**) [39]. We have summarized recent materials hosting a transition-metal-based kagome lattice in Table I.

In this Review, we highlight the kagome R166 (R = rare earths) magnet with RMn$_6$Sn$_6$ being representative, where the quantum interactions within the pristine Mn kagome lattice and with the localized 4$f$ electrons induce ample topological properties[52–54, 77–94]. We first review the experimental visualizations of distinct kagome bands encoded in the unique crystal structure and magnetism. Then we discuss how to tune these topological band structures, especially the Chern gap, by means of manipulating the 4$f$-3$d$ coupling strength and magnetization. The spin texture of the geometrically frustrated Mn kagome lattice can be tuned by external magnetic field, which results in some exotic electromagnetic responses, including exchange-biased effect, topological Hall effect and the emergent inductance. The R166 family, with vast material candidates and highly tunable 3$d$ and 4$f$ magnetic properties, has the potential to be a fertile ground for realizing intriguing kagome physics.

**Crystal structure and magnetism**

RMn$_6$Sn$_6$ compounds belong to the intermetallic RT$_6$E$_6$ (T = transition metals including V, Cr, Mn, Fe, Co, Ni; and E = Ga, Ge, In, Sn, Sb), a grand family containing a few hundred compounds[95–97]. This family has been known for several decades, and many members can be synthesized in the high-quality single crystalline form with a simple flux method[98–101]. One inspiring feature is that this family all possesses a transition-metal-based kagome lattice, either pristine or distorted due to different R arrangements. The R166 compounds can be regarded as a stuffed version of the CoSn structure as shown in Fig. 2**a** [95, 96]. An alternation of Sn-centered Co kagome nets with Sn honeycomb nets creates large hexagonal void spaces in the CoSn-type structure, serving as host to cationic guest atoms. When an R atom enters the cavity as the guest, its chemical pressure pushes the Sn sites within the kagome nets away from the void center, making the two adjacent void spaces along the $c$ axis inaccessible and doubling the $c$ axis of the unit cell. Because the included R atoms can settle either at $z$ coordinate of 0 or 1/2, the ordering patterns create a huge structure variety in the R166 family. The repeat periods range from 2 to 68 CoSn subcells and there are even structurally incommensurate members like the RFe$_6$(Ge/Ga)$_6$ (R =Sc, Tb, Er, Lu) [95–97].

Crystallizing in the HfFe$_6$Ge$_6$-type structure (space group P6/mmm), the RMn$_6$Sn$_6$ sub-family, with R being the heavy rare earth elements (R = Gd - Lu, Y), have the simplest ordering pattern as the R atoms alternatively enter the cavities exactly in the same $z$ coordinate. This simple ordered alteration of R-occupied honeycombs with R-vacant honeycombs was also observed in most of RMn$_6$Ge$_6$ and a few RFe$_6$(Sn/Ge)$_6$ members[95–97, 101–104]. It is suggested that such an alternation makes the R-filled Sn honeycombs appear relatively anionic with the remaining honeycombs being relatively cationic, prompting a coherent electron transfer between the two types of honeycombs to electrostatically stabilize the R cations[96]. This leads to two important consequences: 1, the RMn$_6$Sn$_6$ forms pristine, defect-free Mn kagome layers; 2, the magnetic interaction between the R and Mn moments leads to a variety of magnetic structures and states.

Up to now, only Fe and Mn atoms are known to possess strong magnetic moments in the R166 family[97, 101, 103, 106–108]. The Fe-based siblings retain much of their magnetic properties from the parent FeGe and FeSn compounds, adopting an antiferromagnetic (AFM) order with +-+- stacking of Fe spins along the $c$-axis with the Néel temperature far above room temperature while the R moments order independently at low temperatures. In contrast, the magnetic properties of the Mn-members are rather complex. While the Mn moments are AFM ordered in $RMn_6Ge_6$, the magnetic structures in $RMn_6Sn_6$ vary from strongly anisotropic ferrimagnet (FiM) to AFM, depending on the R elements.

There are three major types of exchange interactions in $RMn_6Sn_6$: the direct Mn-Mn interaction, the indirect $4f$-$3d$ type R-Mn interaction, and RKKY-type R-R interaction. Usually the exchange integrals $|J_{Mn-Mn}| \gg |J_{R-Mn}| \gg |J_{R-R}|$. The intralayer Mn-Mn interaction is the strongest and the Mn kagome lattice tends to form an in-plane FM order. For non-magnetic R = Lu, Sc and Y, the lack of the transferred FM interaction between the Mn layers leads to complex AFM spiral structures[88–92, 98, 102, 109–111]. For magnetic R, the magnetic structure and anisotropy are controlled by the R-Mn interaction. The 12 Mn moments in the hexagonal prism interact with the $4f$ local moment in its center, and the R sublattice orders simultaneously with the Mn sublattice, forming a ferromagnetic (FM) structure for the light rare earths[112–114]. This is due to the fact that the hybridization between Mn-$3d$ and R-$5d$ orbits always favors an antiparallel coupling, while the true exchange coupling between R-$5d$ and local $4f$ orbits is positive[115, 116]. According to the Hund's rule, $J = |L - S|$ for light rare earths, and the compounds favor an FM order. For heavy rare earth R = Gd to Ho members, $J = L + S$, the overall coupling is therefore negative, and R moment orders antiparallel with respect to Mn moment (Fig. 2b), leading to an FiM structure[98, 102, 117]. The R-Mn interaction is too weak for R = Er and Tm, and these two members fall back to the AFM structure. For detailed discussions of the magnetic structures, one may refer to REF. [101]. It should be noted that $YbMn_6Sn_6$ poses an exception to the above regulation because the Yb ions are divalent. Alkaline (A = Li) and alkaline earth (AE = Mg and Ca) $A(AE)Mn_6Sn_6$, together with $YbMn_6Sn_6$, are all FM ordered[118–120]. The change of the valence electron concentration and the modification of the chemical bonding scheme in the Mn-(Sn-R)-Mn slab due to the less electronic affinities of +I and +II ions may account for the difference[119].

Figure 2c shows that the magnetic anisotropy of $RMn_6Sn_6$ varies from easy-$ab$-plane for Gd to a conical magnetic structure for R = Dy and Ho [84, 98, 101, 102, 117]. Interestingly, $TbMn_6Sn_6$ (hereafter Tb166 for short) is the only one possessing an easy-$c$-axis anisotropy. It undergoes two successive magnetic transitions, paramagnetic to easy-plane FiM order at $T_C$ = 420 K and easy-plane to perfect out-of-plane FM order at $T_{C2}$ = 313 K. The out-of-plane FM Mn kagome lattice is stable over a wide range in its phase diagram (Fig. 2d), which is crucial to support the fully spin-polarized Dirac fermions according to recent experimental works[42, 52, 53]. In a tight binding model for kagome lattice, the inclusion of SOC will further open an energy gap (Chern gap) at the Dirac cone with a non-zero Chern number, which effectively realizes the spinless Haldane model. This simple model suggests a route for realizing the anomalous quantum Hall effect at high temperatures[26, 27, 33]. The unique out-of-plane FM Mn moments in Tb166 at room temperature have motivated wide research interests for exploring the topological electrons in the kagome lattice[52–54, 80–87].

**Spectrum of Chern-Gapped Dirac fermion**

It is the scanning tunneling microscopy (STM) measurement on Tb166 that detected the first experimental evidence for the existence of the Chern-gapped Dirac fermion in kagome lattice. Over the past decade, STM has become a powerful tool to probe and discover emergent topological matter. It utilizes the quantum tunnelling principle[121–123] to probe the surface morphology and the local density of states of materials with atomic-scale precision and (sub-) meV energy resolution, and is compatible with magnetic field tunability. For in-depth reviews of the STM technology and its condensed matter applications, one may refer to REF.[105, 123].

Due to the Tb interception in Tb166, the interlayer distance between the TbSn layer and Mn layer is large and the crystals tend to cleave between these planes[52]. It was therefore possible to obtain the atomically flat Mn kagome lattice as shown in Fig. 3**a**. Unlike topographies of the kagome lattice in other kagome magnets which show various atomic defects[43, 61, 124–126], there is no detectable defect over a large field of view in Tb166, offering an unprecedented opportunity to explore the intrinsic topological quantum properties. It was found that the low-energy tunneling spectrum of the Mn kagome layer exhibits unique Landau quantization under magnetic field, with several key features (Fig. 3**b**): Firstly, the zero-field peak is shifted linearly to lower energy with increasing field, which indicates magnetic polarization with a Zeeman term ($\Delta E = \mu_B B/2$, $\mu_B$ is the Bohr magneton) rather than splitting. Secondly, other Landau levels shift with a square root-like field dependence below this state, consistent with Dirac-like fermions featuring the spectrum $\varepsilon_n \sim \sqrt{nB}$. Thirdly, another intense state emerges above the zero-field peak under field and shifts in parallel with it, defining a gap $\Delta$ in between. All the observations highlight the existence of spin-polarized Dirac electrons with a large Chern gap in Tb166 [52].

The observed Landau level spectrum can be well depicted by the gapped Dirac model

$$E_n = E_D \pm \sqrt{(\Delta/2)^2 + 2|n|e\hbar v^2 B} - g\mu_B B/2, \qquad (1)$$

where the Dirac cone energy $E_D$ = 130 meV, the gap $\Delta$ = 34 meV, the Dirac velocity $v$ = 4.2 × 10$^5$ m/s and Landé g-factor g = 52. Based on these parameters extracted from the Landau level spectrum, the band dispersion was obtained, which showed agreements with occupied bands observed by angle-resolved photoemission spectroscopy (ARPES) data (Fig. 3**d**). Precisely in the gap, a pronounced step edge state was also observed (Fig. 3**e**), verifying the existence of a chiral edge state within the Chern gap. These spectroscopic observations point to the realization of a quantum-limit Chern phase in Tb166.

**Transport properties of Chern-gapped Dirac fermion**

As the Chern-gapped Dirac fermion lies close to the Fermi level in Tb166, a direct question is whether the relativistic particle manifests pronounced anomalous electromagnetic response. Experiments revealed two keynote transport features of the massive Dirac fermion, the Shubnikov-de Haas quantum oscillations (SdH QOs) and the anomalous transverse transport[54]. Strong quantum oscillations with one dominant frequency (96 T) were observed in the thermoelectric and electric transport measurements at low temperatures (Fig.

**4a**). Standard analysis of the oscillatory patterns[127, 128] revealed that this orbit is from a hole-like band with a high Fermi velocity $v = 4.2 \times 10^5$ m/s and the Fermi energy $E_D$ was estimated to be around 140 meV, in remarkable accordance with the hole branch of the Dirac dispersion observed in the spectroscopic experiments[52]. The oscillatory frequency and cyclotron mass follow the angle dependence of $1/cos\theta$ (Fig. **4c**) for the gapped Dirac dispersion[129]. The Berry phase, which was subtracted from oscillatory peak positions (Fig. **4b**), also suggests a nontrivial band topology[130, 131]. All these demonstrate that the quasi-2D electronic pocket from Mn kagome lattice plays a vital role in transport.

The anomalous Hall effect (AHE), anomalous Nernst effect (ANE) and anomalous thermal Hall effect (ATHE) represent closely-related, off-diagonal electric, thermoelectric and thermal signals in the absence of external magnetic field, respectively. Stemming from the electron's anomalous velocity endowed by the Berry curvature field in the $k$ space[1], they are important fingerprints of the topological band structures residing near the Fermi energy, and can be formulated in terms of Berry curvature $\Omega_z(k)$ with the general form[132–134],

$$\mathcal{C}_n = -\frac{1}{\hbar}\int d\varepsilon \int_{BZ}\frac{d\boldsymbol{k}}{(2\pi)^3}\left(\frac{\varepsilon-\mu}{k_BT}\right)^n \frac{\partial f(\varepsilon-\mu)}{\partial \varepsilon}\sum_{\epsilon_n<\varepsilon}\Omega_z^n(\boldsymbol{k}). \qquad (2)$$

The anomalous Hall conductivity (AHC) $\sigma_{xy}^A$, the thermoelectric Hall conductivity $\alpha_{xy}^A$ and the anomalous thermal Hall conductivity $\kappa_{xy}^A$ can be expressed in terms of $\mathcal{C}_n$,

$$\sigma_{xy}^A = e^2 \mathcal{C}_0, \qquad \alpha_{xy}^A = k_B e \mathcal{C}_1, \qquad \kappa_{xy}^A = k_B^2 T \mathcal{C}_2. \qquad (3)$$

It was found that Tb166 presents large anomalous transverse coefficients from intrinsic Berry curvature contributions (Fig. **4d-f**), with magnitude comparable to other topological magnets[47, 134–138]. For the Chern gapped Dirac bands with Chern number $C$, $\sigma_{xy}^{int}$ is calculated[139] as

$$\sigma_{xy}^{int} = C \cdot \frac{\Delta/2}{\sqrt{(\Delta/2)^2 + \hbar^2 k_F^2 v_D^2}}, \qquad (4)$$

with band parameters determined from STM[52] and transport[54]. By scaling the AHC $\sigma_{xy}^A$ versus conductivity $\sigma_{xx}$ (Fig. **4h**) [140], an intrinsic contribution of $\sigma_{xy}^{int} \sim 0.13$ e$^2$/h per kagome layer was extracted[52]. The Chern quantum number was therefore evaluated to be $C = 1$.

The scaling relations between $\sigma_{xy}^A$, $\alpha_{xy}^A$ and $\kappa_{xy}^A$ demonstrate the topological charge-entropy transport of the Dirac fermion. A conventional version in solids is the widely known Mott formula and Wiedemann-Franz law[141, 142], which associate the charge and entropy of the electron via $\boldsymbol{\alpha} = \frac{\pi^2}{3}\frac{k_B^2 T}{e}\frac{\partial \boldsymbol{\sigma}}{\partial \varepsilon}|_{\varepsilon_F}$, $\boldsymbol{\kappa} = L_0 \boldsymbol{\sigma} T$, where $L_0 = 2.44 \times 10^{-8}$ V$^2$/K$^2$ is the Sommerfeld value, $\boldsymbol{\sigma}$, $\boldsymbol{\alpha}$ and $\boldsymbol{\kappa}$ are the electric, thermoelectric and thermal conductivity tensor, respectively. In the anomalous transverse transport in Tb166, however, their relevances are modified into (Fig. **4i-j**)

$$\alpha_{xy}^A/\sigma_{xy}^A = \frac{\pi^2}{3}\frac{k_B}{e}\frac{k_B T}{E_D} \qquad (5)$$

$$\kappa_{xy}^A/\sigma_{xy}^A = L_0 T\left[1 + \eta\left(\frac{k_B T}{E_D}\right)^2\right], \eta = \frac{7\pi^2}{5} \qquad (6)$$

The former equation unearths the physics behind the empirical $k_B/e$ threshold[136, 138] for $\alpha_{xy}^A/\sigma_{xy}^A$ in the kagome Chern magnet, while the latter presents a living example that the Wiedemann-Franz law can be violated by a mismatch between the thermal and electrical summations on the singularity of Chern gap's Berry curvature in Eq.(2). The fine consistency between experiments and the 2D Dirac model is probably due to the fact that all other Berry curvature contributions from the 3D dispersive bands are washed out in the ratio between these integrations because they are relatively smoother. Shortly after this work, a theoretical work by Wang *et al.* confirmed this interpretation using a heat-bath method, and further suggested that the upturn at 100 K in $\alpha_{xy}^A/\sigma_{xy}^A$ (Fig. 4**i**) may be a manifestation of inelastic dissipation[143].

**Topological band engineering**

The discovery of the Chern-gapped Dirac fermion in Tb166 motivated the extended investigation of the topological band in other R166. For local-moment-bearing R, the magnetic structure of $RMn_6Sn_6$ varies from FiM easy-plane (R = Gd), easy-axis (R = Tb), easy-cone (R = Dy and Ho) to AFM (R = Er and Tm)[98, 101, 102, 117]. The magnetic anisotropy is determined by the crystalline electric field on the R elements while the repulsive crystal-field interaction with Mn atoms stabilizes the different spin structures of R [84]. As the massive Dirac fermion is hosted in the out-of-plane, FM Mn-based kagome lattice in Tb166, it is intuitive to expect that the topological band is similarly hosted in other $RMn_6Sn_6$ when an external field polarizes the Mn spins along the *c* axis.

Magneto-transport on those field-polarized FiM $RMn_6Sn_6$ (R = Gd to Er) detected SdH QOs with a close and small oscillatory frequency and dominant intrinsic AHE[53]. Combined with spectroscopic experiments[52], it is suggested that the electronic structures of $RMn_6Sn_6$ resemble each other and the Chern gapped Dirac fermion is generally hosted in FM Mn kagome lattice of $RMn_6Sn_6$ (R = Gd to Er) when the Mn spins are polarized along the *c* axis. Applying Eq. (4) and the constraints from QOs, it is found that two defining parameters of the massive Dirac band, the gap size $\Delta$ and the cone energy $E_D$ evolve continuously against the de Gennes factor ($dG = (g_J - 1)^2 J(J + 1)$, where $g_J$ is the Landé factor and $J$ is the total angular momentum of the $R^{3+}$ ion), following a linear and square root relation, respectively (Fig. 5**b**). Based on these observations, a possible Chern gap opening mechanism was proposed by introducing electron hopping between R and Mn atoms[53]. The gap opening of the 2D Dirac cone has also been revealed in a theoretical study[84].

The studies on $RMn_6Sn_6$ unveiled that the R atoms play an important role in the topological band structures while the substitution of the R atoms can effectively tune the topological band. Substituting Gd with Tb in alloy $Tb_xGd_{1-x}Mn_6Sn_6$ smoothly tunes the magnetic anisotropy from in-plane to out-of-plane FiM without introducing impurity on Mn kagome layer[144, 146]. Due to the in-plane FM order of the Mn kagome lattice, the Chern gap in the parental GdMn$_6$Sn$_6$ should vanish, which is verified in first-principle calculations[84, 85, 144, 147] where the Dirac point formed by electron-like band 40 meV below the Fermi level possesses a neglecting SOC gap (<0.5 meV). It is suggested such a small SOC gap is due to the remarkable cooperative interplay between the Kane-Mele SOC, low site symmetry and in-plane magnetic order[144]. The ARPES experiments found that upon 20% substitutional Tb doping (Tb$_{0.2}$Gd$_{0.8}$Mn$_6$Sn$_6$), a significant band gap (25 meV) opens at this Dirac cone (Fig. 5c) as long as the magnetic reorientation to the out-of-plane easy axis occurs[144].

As the topological band structures are determined by the magnetic structure of R166, it can be effectively modulated by an external magnetic field as well. Currently relevant research mainly focuses on YMn$_6$Sn$_6$ where the Mn spin is in an incommensurate flat spiral configuration and can be gradually aligned by an external field[89, 91, 109, 148]. In an in-plane magnetic field, the continuously varying spin textures may change the Fermi surface topology, causing a Lifshitz transition and a sharp reduction of the magnetoresistance up to 45%[148]. Under an out-of-plane magnetic field, STM experiments have also identified the manipulation of the massive Dirac cone near the Fermi level in YMn$_6$Sn$_6$[145]. By subtracting the band dispersions from quasiparticle interference (QPI) data, it was found that in addition to an overall Zeeman shift of the massive Dirac cone with increasing field, the band bottom of the Dirac cone is significantly rounded (Fig. 5d) due to a momentum-dependent $g$ factor peaking around the Dirac point (Fig. 5e). Several mechanisms have been proposed to explain this evolving $g$ factor, including the large orbital magnetic moments primarily localized near the Dirac points, the spin canting induced Dirac gap enhancing, as well as the evolving magnetic exchange coupling[145].

**Hidden order, spin dynamics and exchange-bias**

So far we have focused mainly on the band topology and its direct interplay with the magnetic ground states of R166. There is another perspective from the viewpoint of spin dynamics: the frustration and fluctuation. The kagome lattice naturally favors geometrically frustrated interplanar exchange interactions, which promise it as an ideal candidate hosting the spin liquid phase[40, 41]. Even in a long-range ordered magnetic state, some hidden magnetic orders may be induced by the geometric frustration and thermal fluctuation. For instance, the Co-based kagome metal Co$_3$Sn$_2$S$_2$ was assumed to be a simple easy-*c*-axis FM below T$_C$ = 175 K[45, 50, 125, 137, 149], yet recent studies showed that there exist unconventional magnetic phase transitions at around 130 K[150–155]. The μSR experiment, which serves as an extremely sensitive local probe for detecting microscopic details of the static magnetic order, ordered magnetic volume fraction, and magnetic fluctuations, suggested the magnetic phase separation where competing in-plane AFM state coexist with the FM state below T$_C$[151]. Some studies suggested the existence of unconventional states such as spin glass and local symmetry breaking[154, 156]. The scanning Kerr microscopy measurements, on the other hand, suggested that the transition at 130 K may be embedded inside the 2D magnetic domain wall, in which the magnetization texture changes from continuous rotation to unidirectional variation due to a large anisotropy energy to magnetostatic energy ratio[153]. Regardless of the diverse interpretations of the microscopic origin, the magnetic phase separation and hidden orders in $Co_3Sn_2S_2$ lead to a pronounced exchange-bias effect in the field-dependent isothermals below the $T_C$ [156, 157]. The exchange-bias effect has potential in magnetic memory technologies as the hysteresis loops of magnetization, AHE, and the ANE, ATHE shift along the applied magnetic field axis[158, 159].

Relevant research revealed that the spin dynamics and unconventional magnetic order play an important role in Tb166 which was thought to be a simple ferrimagnet with two transitions, from PM to easy-plane FM at around 420 K and to easy-$c$-axis FM at 310 K [98, 102, 117]. A recent study by Mielke *et al.* found a huge hysteresis between zero-field-cool ($\chi_{ZFC}$) and field-cool ($\chi_{FC}$) susceptibility below $T^*_{C1} \simeq 120$ K when the applied magnetic field is very small (5 mT) [82], as shown in Fig. 6**a**. The zero-field-cooling $\chi_{ZFC}$ further settles into a negligibly small diamagnetic plateau below $T_{C1} = 20$ K. Employing the μSR technique, the authors found a fast depolarization of the implanted muons in a wide temperature range, a direct evidence of the involvement of slow magnetic fluctuations in the out-of-plane FiM state of Tb166. The magnetic fluctuations slow down quickly below $T^*_{C1}$ as indicated by the large increase of the depolarization rate $\lambda_L$ of the μSR signals (Fig. 6**b**), and finally freeze into quasi-static patches with ideal out-of-plane order below $T_{C1}$. $T^*_{C1}$ and $T_{C1}$ therefore correspond to two critical fluctuation strengths. It was suggested that the spin reorientation transition at 310 K also arises from the competition between the stronger fluctuating Tb moments and the less fluctuating Mn moments[83].

The exchange-biased anomalous transverse effect is also observed in Tb166 [80]. Zhang *et al.* found that when the sample is field cooled before measurements, asymmetric magnetization hysteresis loops could be observed at low temperatures (Fig. 6**c**). The exchange-biased field ($H_{EB}$), which is defined as the shift of the center of hysteresis loops away from the origin of the field axis, increased monotonically below 200 K (Fig. 6**d**). Similar features were also observed in AHE, ANE and ATHE. The exact mechanism behind the exchange bias effect remains unclear, but there is some evidence showing its relevance with the magnetic fluctuations[82]. It was speculated that in Tb166, as the magnetic fluctuation begins to freeze when the temperature decreases, the competition of interlayer couplings between Mn spins, together with the FiM Tb-Mn interlayer coupling, may lead to the formation of the cluster spin-glass state with AFM domains embedded in and coexisting with the bulk FM phase of Mn spins[80]. As a result, the interface between AFM domains and FM phase of Mn spins gives rise to the exchange-bias features. Because the magnetization of Tb166 is comparable to $H_{EB}$, this exchange-biased effect might facilitate the integration of Nernst thermopile overcoming the stray field interference[160]. Giving that the complicated spin dynamics and magnetic interactions for the different 4$f$ local moments and Mn spins may be a shared feature in $RMn_6Sn_6$, the relevant research on the other R166 members will be a vast open field.

**Topological Hall effect and emergent inductance**

The complicated spin configurations of R166 give rise to another interesting topological phenomenon which is related to the real-space Berry curvature field. When conduction electrons flow with their spin direction aligned along an underlying spin structure, they can acquire a Berry phase[1, 161]. This also acts as an effective electromagnetic field, termed the emergent electromagnetic field, which, in the continuum limit, can be described as[1, 162]

$$b_i = \frac{h}{8\pi e}\epsilon_{ijk}\, \boldsymbol{n} \cdot (\partial_j \boldsymbol{n} \times \partial_k \boldsymbol{n}), \tag{7}$$

$$e_i = \frac{h}{2\pi e}\, \boldsymbol{n} \cdot (\partial_i \boldsymbol{n} \times \partial_t \boldsymbol{n}), \tag{8}$$

where $b_i$ and $e_i$ are the emergent magnetic and electric fields, respectively, $\boldsymbol{n}$ is a unit vector parallel to the spin direction, and $\epsilon_{ijk}$ is the Levi-Civita symbol. The emergent magnetic field $b_i$ arises from non-coplanar spin structures (for example, the magnetic skyrmion[162, 163]) and is proportional to the solid angle subtended by $\boldsymbol{n}$. This can produce an additional term in the Hall response, dubbed as the topological Hall effect (THE) [164–166].

The effect of an emergent electric field $e_i$ is less experimentally studied. It is related to the dynamics of spin structures and proportional to the solid angles swept out by $\boldsymbol{n}(t)$. As the canonically conjugated spin position ($X$) and the angle tilt $\phi$ from the helical plane change simultaneously with time, the motion of non-collinear spin structures can also induce nonzero $e_i$ which is proportional to the time-derivative of $\phi$, i.e. $\partial_t \phi$ (Fig. 7e) [167, 168]. When a non-collinear spin structure is driven by a current parallel to the magnetic modulation vector due to spin-transfer or spin-orbit torque, the emergent electric field is predicted to produce an inductive voltage[168], which was recently observed in the helimagnet $Gd_3Ru_4Al_{12}$ [169].

The emergent electromagnetic field may exist in the R166 family due to the vast variety of magnetic states. The in-plane helical spiral spin configuration in $YMn_6Sn_6$ has attracted particular interest. The crystal structure of $YMn_6Sn_6$ consists of kagome slabs [$Mn_3Sn$] separated by two inequivalent $Sn_3$ and $Sn_2Y$ slabs, i.e., [$Mn_3Sn$][$Sn_3$][$Mn_3Sn$][$Sn_2Y$]. All Mn planes and in-plane nearest-neighbor Mn-Mn bonds are crystallographically equivalent, while the interplanar Mn-Mn bonds along $c$ have an FM exchange interaction across the $Sn_3$ layers but AFM across the $Sn_2Y$ layers. The frustration due to the second-neighbor interaction across an intermediate [$Mn_3Sn$] slab results in complex magnetic behaviors[88, 89, 91, 92, 110]. Neutron scattering experiments show that below $T_N \approx 345$ K, a commensurate collinear AFM structure first forms with the propagation vector k = (0, 0, 0.5). Upon cooling, an incommensurate phase quickly appears, which coexists with the commensurate phase in a narrow temperature range and becomes the only phase below 300 K [88, 110]. This incommensurate state has been reported to have two nearly equal wave vectors[110], which can be described as a "double flat spiral" [111] (Fig. 7b, left panel). In the existence of an applied magnetic field in the $ab$ plane, the phase diagram of $YMn_6Sn_6$ is composed of several complicated magnetic structures (Fig. 7a). With increasing in-plane field, the spin texture changes from distorted spiral (DS), to transverse conical spiral (TCS, with a $c$-axis component), fan-like (FL), and eventually field polarized FM (FF) (Fig. 7b) [89, 91].

It was found that the spin texture changes in $YMn_6Sn_6$ can induce multiple topological transitions[148]. Significant THE was observed in the TCS state[89, 91] (Fig. 7c). Here the total Hall resistivity $\rho_H$ can be described as $\rho_H = \rho_H^N + \rho_H^A + \rho_H^T = R_0 B + 4\pi R_s M + \rho_H^T$. By subtracting the normal and anomalous Hall resistivity, a large $\rho_H^T$ was observed at around 245 K and a magnetic field of 4 T (Fig. 7**d**) [89]. This THE was explained as the result of the nonzero spin chirality of the TCS state[91], and thermal fluctuation may also play a role in stabilizing a new nematic chirality[89].

Such unusual spin textures can generate the emergent inductance in YMn$_6$Sn$_6$ as well[170]. Kitaori *et al.* fabricated micro-scale devices of YMn$_6$Sn$_6$ and measured the imaginary part of the resistance. It was found that when the electric current is applied along the *c* direction, an inductance up to a few μH is observed in both the DS and TCS states (Fig. 7**f**), whereas in the FF state the inductance vanishes as expected. The observed inductance value $L$ and its sign are found to vary to a large extent, depending not only on the spin-helix structure controlled by temperature and magnetic field but also on the applied current density. This may provide a step towards realizing microscale quantum inductors. The emergent inductance in Gd$_3$Ru$_4$Al$_{12}$ and YMn$_6$Sn$_6$ has inspired theoretical researches taking account of the spin relaxation, sample disorder and collective modes[171, 172].

**Outlook**

In this review, we have mainly discussed two topological aspects of the R166 family: the topological band structures in the reciprocal space, and the topology of spin structures in the real space. Up to now, most studies were carried out on two members, TbMn$_6$Sn$_6$ [52–54, 80–87] and YMn$_6$Sn$_6$ [88–92, 111, 145], which represent the two ends of the magnetic exchange coupling strength. In addition to the shared band structures of a kagome lattice, the former features a Chern-gapped Dirac fermion due to the strong ferromagnetism in the kagome lattice which is induced by the 4*f*-3*d* coupling, while the latter, with no 4*f* moments, shows complex transport properties closely related to the helical spin texture. Whether there exist intertwined topological phenomena in other Mn-based R166 members is still an open question.

Recently, the strongly correlated nature of the electrons in RMn$_6$Sn$_6$ was suggested by the DFT calculations and infrared spectroscopy experiments. The DFT calculations have reproduced most of the band dispersions observed in ARPES experiments, but they expect the 2D Dirac cone at the $K$ point to be about 0.7 eV above the Fermi level[52, 78, 83–85, 144, 147]. In contrast, STM, ARPES and transport experiments all confirmed the existence of 2D Dirac dispersion just above the Fermi level (around 0.13 eV for TbMn$_6$Sn$_6$ and GdMn$_6$Sn$_6$) [52, 53, 144]. As this Dirac cone is the one with the highest Fermi velocity and minimal $k_z$ dispersion[52, 84], its exact location is vital for the interpretation of transport phenomena. A possible explanation for the discrepancy is the correlation effect embedded in the Mn kagome lattice. RMn$_6$Sn$_6$ is a good metal, and Mn electrons are on the itinerant side. Yet these *d* electrons are still considerably, though not strongly localized, and it is hard to capture in static methods such as LDA+U or hybrid functionals. The dynamical field theory (DMFT) also faces severe problems in RMn$_6$Sn$_6$ where long-range correlations are expected and hybridization with Sn is crucial[84]. It has been suggested that the ratio of the spectral weight of the mobile carriers from the optical experiment and the DFT calculation (SW$_{Drude}$/SW$_{band}$) can be used as a gauge of electronic correlation strength[173, 174]. Here SW$_{Drude}$/SW$_{band}$ is close to 1 for uncorrelated materials, while the ratio becomes 0 for Mott insulators showing the strongest correlation. The study of Wenzel *et al.* shows that the ratio for RMn$_6$Sn$_6$ is around 0.2, comparable to the cuprate superconductors La$_{2-x}$Sr$_x$CuO$_4$ [87, 175]. The result indicates much stronger correlations in comparison with the AV$_3$Sb$_5$ series and other kagome metals reported to date. Lee *et al.* [84] also found that when considering correlations in the LDA+U method, the 2D dispersion in Tb166 can be greatly shifted downwards for 0.4 eV.

Recently, the nonmagnetic T-based R166 including RV$_6$Sn$_6$, RCo$_6$Ge$_6$ and RCr$_6$Ge$_6$ has also drawn attention. The RV$_6$Sn$_6$ group manifests characteristic kagome band structures including the Dirac cone, saddle point, and flat bands[67–69, 176–180]. One interesting observation is a structural modulation with wave-vector (1/3, 1/3, 1/3) in ScV$_6$Sn$_6$ [69–71], suggesting that the charge order might be a common tendency in the partly filled *d*-orbital kagome systems[181–184]. A structure distortion doubling unit cell's *c* axis is also found in YbCo$_6$Ge$_6$ [185]. These observations are reminiscent of the kagome AV$_3$Sb$_5$ family hosting (1/2, 1/2, 1/2) or (1/2, 1/2, 1/4) CDW phases and unconventional superconductivity, which are closely related to the van Hove singularities near the Fermi level[25, 32, 34–39]. Considering the wide variety of structures and magnetisms, whether the correlation related many-body effects like the unconventional superconductivity can be realized in the R166 family is a deserving task in future.

**Tables**

**Table I Summary of the recently studied kagome magnets.**

| Materials | Synthesis | Property | Magnetism | $T_{C/N}/T_{SC}$(K) | Refs. |
|---|---|---|---|---|---|
| $ZnCu_3(OH)_6Cl_2$ | hydrothermal | quantum spin liquid | no LRO observed | – | 57,58 |
| $Cs_2LiMn_3F_{12}$ | solid reaction | Chern phase | in-plane FM (calc.) | – | 27,59 |
| $Mn_3Sn$ | Bridgman, Czochralski | Weyl semimetal | non-colinear AFM | 420 | 46,47 |
| $Mn_3Ge$ | Bridgman, Czochralski | Weyl semimetal | non-colinear AFM | 380 | 48,60 |
| $Co_3Sn_2S_2$ | flux, CVT | Weyl semimetal, flat band | out-of-plane FM | 177 | 43,49,50 |
| $Fe_3Sn_2$ | CVT | Weyl semimetal, flat band | in-plane FM | 670 | 42,51,61 |
| FeSn | flux | flat band | AFM | 365 | 62 |
| CoSn | flux | flat band | PM | – | 63 |
| FeGe | CVT | CDW | AFM | 410 | 64,65 |
| $YCr_6Ge_6$ | flux | flat band | PM | – | 66 |
| $TbMn_6Sn_6$ | flux | Chern phase | out-of-plane FiM | 423 | 52-54 |
| $GdV_6Sn_6$ | flux | vHs | AFM | 5 | 67,68 |
| $ScV_6Sn_6$ | flux | CDW | PM | – | 69-71 |
| $KV_3Sb_5$ | flux | CDW, vHs | SC | 0.93 | 32,55 |
| $RbV_3Sb_5$ | flux | CDW, vHs | SC | 0.92 | 72,73 |
| $CsV_3Sb_5$ | flux | CDW, vHs | SC | 2.5 | 39,56,74 |
| $LaRu_3Si_2$ | Czochralski | – | SC | 6.5 | 75 |
| $CeRu_2$ | Czochralski | – | SC | 6.5 | 76 |

Only representative compounds in R166 family are listed. $T_{C/N}$, Curie/Neel temperature; $T_{SC}$, superconducting transition temperature; LRO, long-range order; FM, ferromagnetism; AFM, antiferromagnetism; PM, paramagnetism; FiM, ferrimagnetism; SC, superconductor; CVT, chemical vapor transport; CDW, charge density wave; vHs, van Hove singularity.

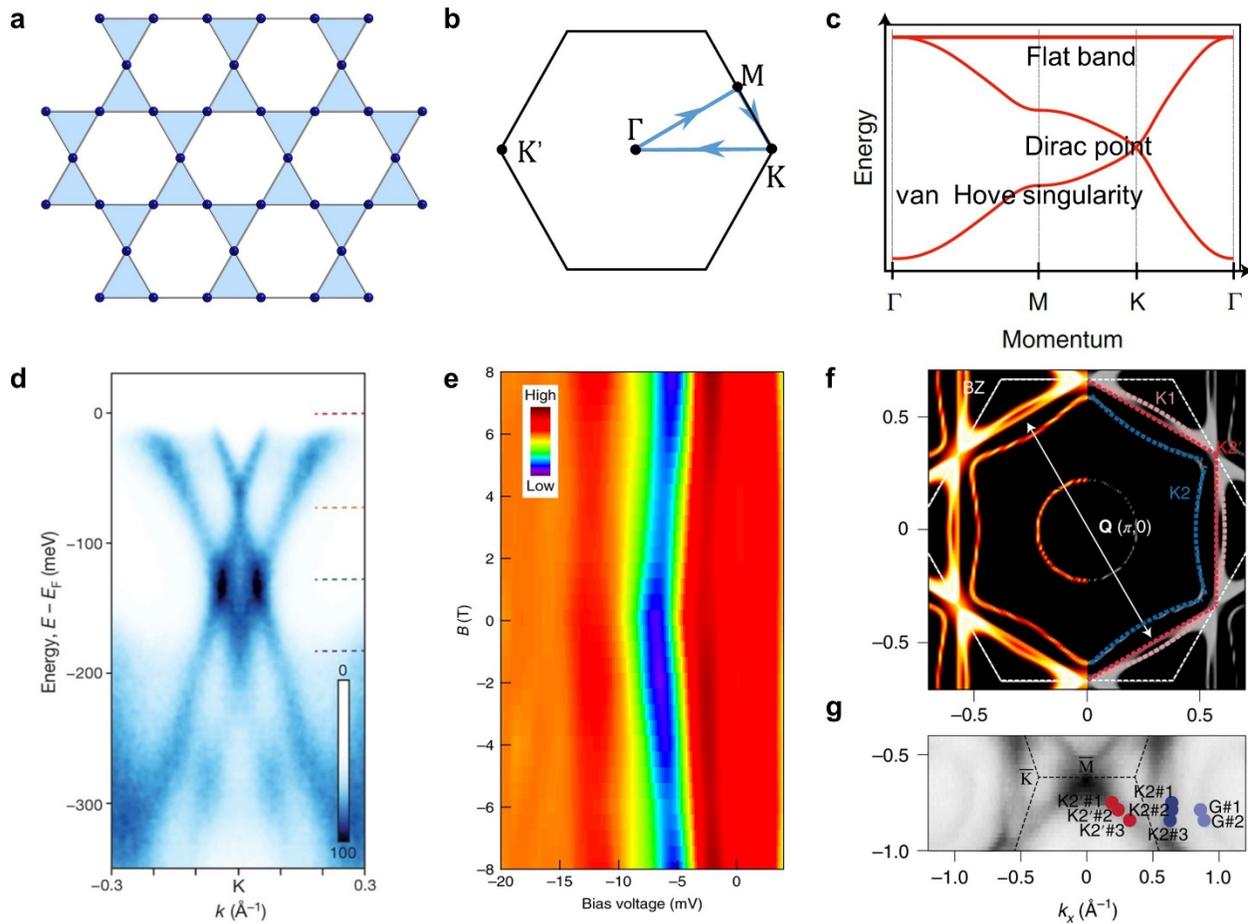

**Figure 1 The kagome lattice and its topological band structure. a-b,** The structure of the kagome lattice and its first Brillouin zone. **c,** Tight-binding band structure of a kagome lattice, showing the existence of the Dirac point at $K$, van Hove singularity at $M$, and the flat band across the Brillouin zone. **d,** Angle resolved photoemission spectroscopy (ARPES) data measured in the kagome metal $Fe_3Sn_2$, showing multiple Dirac cones around the Brillouin zone cone $K$. **e,** The flat band state viewed by scanning tunnelling experiments in the kagome magnet $Co_3Sn_2S_2$, which can be tuned via an out-of-plane vector field. **f,** Calculated Fermi energy surface of $CsV_3Sb_5$ at $k_z = 0.42\pi/c$ where the van Hove singularities from $K2$ and $K1$ bands cross $E_F$. The double-headed arrows indicate the nesting vector compatible with the $2 \times 2$ charge order. **g,** Band- and momentum-resolved charge order gaps of $CsV_3Sb_5$. Panel **c** adapted from REF.[25]. Panel **d** adapted from REF.[42]. Panel **e** adapted from REF.[43]. Panel **f** and **g** adapted from REF.[39].

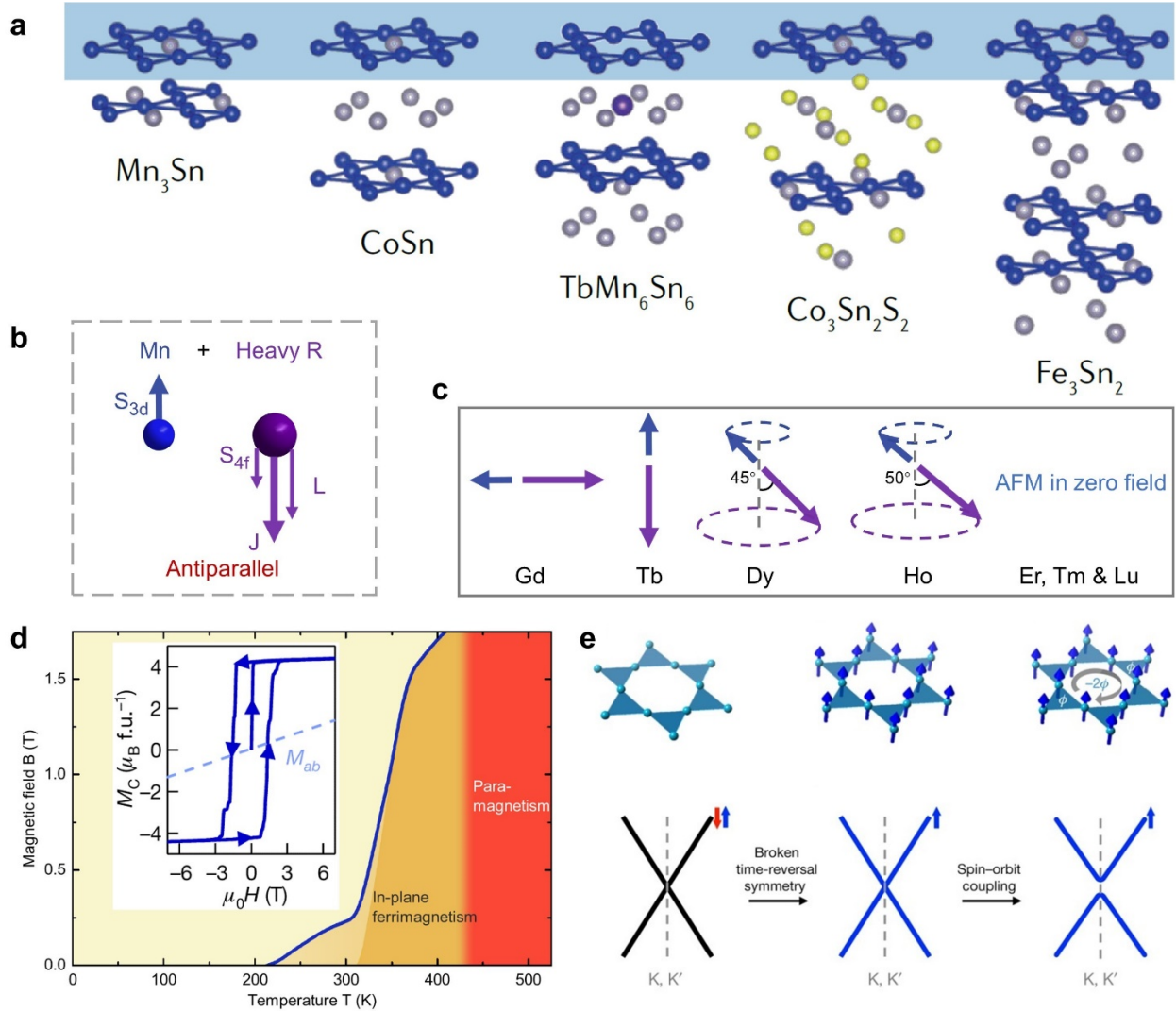

**Figure 2 Structural and magnetic properties of $RMn_6Sn_6$. a,** The crystal structures of representative kagome metals. Unlike $Mn_3Sn$, $CoSn$, $Co_3Sn_2S_2$ and $Fe_3Sn_2$, $RMn_6Sn_6$ features a pristine, A-A stacking kagome lattice, namely no extra Sn atoms in the kagome layer. **b,** The inclusion of rare-earth element in $RMn_6Sn_6$ offers an opportunity of engineering the kagome magnetization. For heavy rare earth elements, the exchange interaction usually forces an antiparallel coupling between $3d$ moments within the transition-metal-based kagome layer and the anisotropic $4f$ moments. **c,** Magnetic structures of $RMn_6Sn_6$ in zero field. Blue and purple arrows represent the direction of Mn and R moments, respectively. **d,** Magnetic phase diagram of $TbMn_6Sn_6$ when the magnetic field is applied along the crystallographic $c$ axis. There are three main regions, including the out-of-plane ferrimagnetic state where Chern-gapped Dirac states are supported, in-plane ferrimagnetic state, and high temperature paramagnetic state. Inset shows the magnetization at 2 K. **e,** The degeneracy of the Dirac dispersion in the kagome lattice can be lifted by the inclusion of magnetization, and the spin-polarized band can further acquire a Chern gap when SOC is included. Panel **a** adapted from REF.[105]. Panel **c** adapted from REF.[53]. Panel **d** adapted from REFs.[52,54]. Panel **e** adapted from REF.[42].

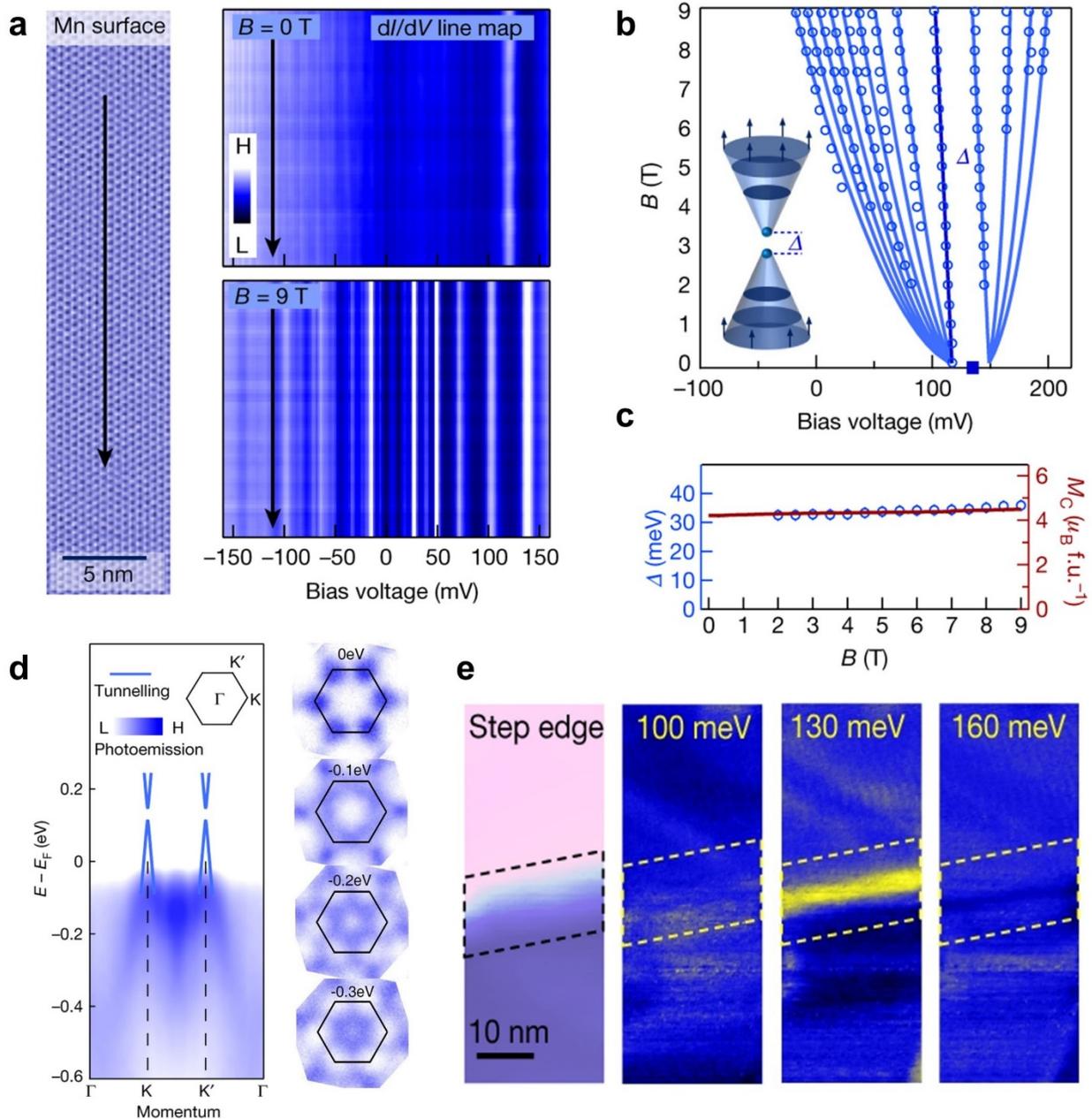

**Figure 3 Spectroscopic observation of the Chern Dirac gap in TbMn$_6$Sn$_6$. a,** Topography of the Mn kagome lattice in TbMn$_6$Sn$_6$ and corresponding dI/dV line maps taken at B = 0 T and 9 T. The intense modulation shown in 9 T data is associated with Landau quantization. **b,** Fitting the Landau fan data with the spin-polarized and Chern gapped Dirac dispersion. **c,** Dirac gap size and out-of-plane magnetization as a function of the magnetic field. **d,** ARPES data taken at a phonon energy of 120 eV, showing the band dispersion along high symmetry lines (left) and the constant energy maps (right). **e,** dI/dV maps taken at different energies across a step edge (left). The map taken within the Chern gap energy (130 meV) shows a pronounced step-edge state. Adapted from REF.[52].

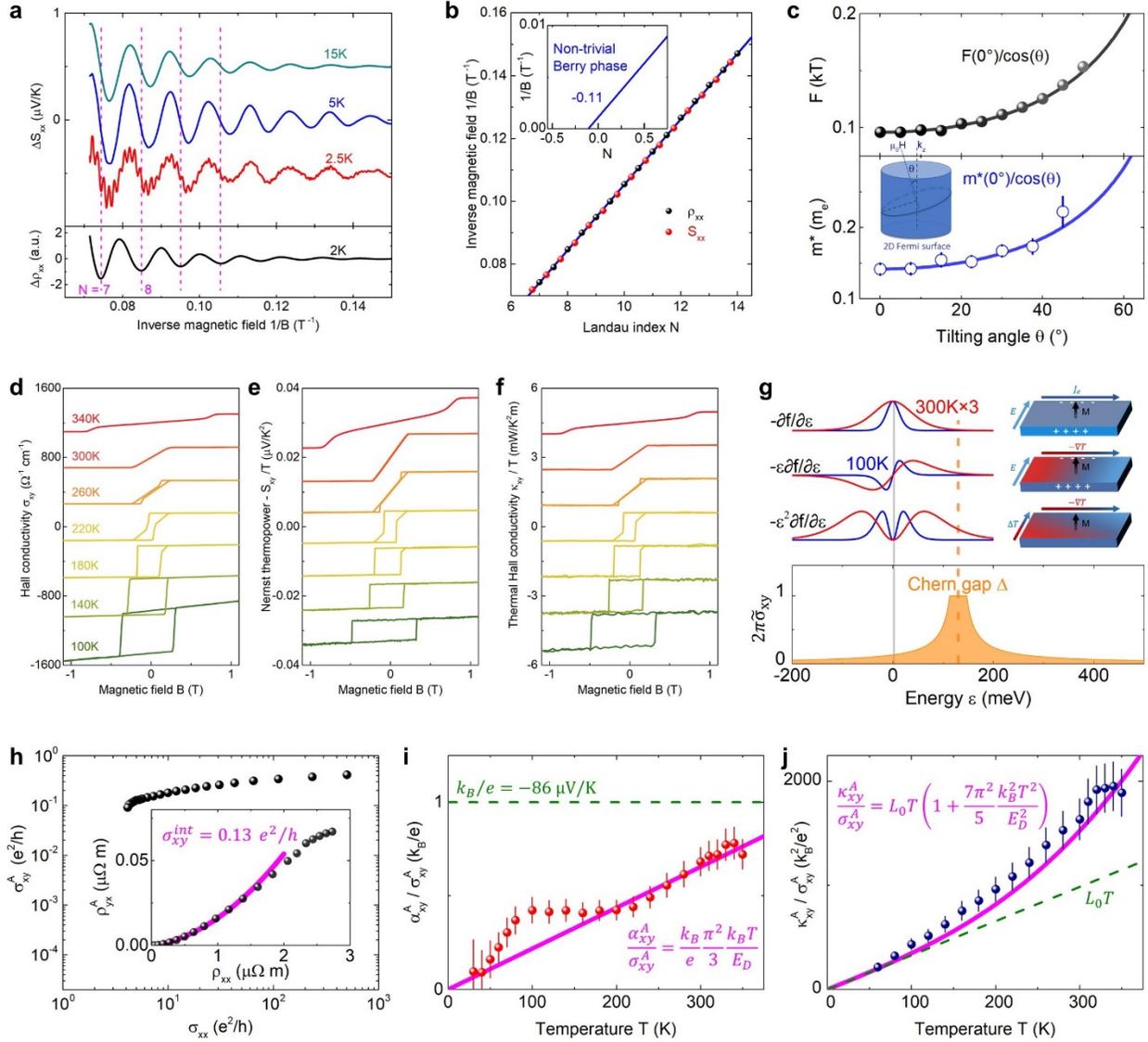

**Figure 4 Transport visualization of the Chern Dirac fermion in TbMn$_6$Sn$_6$. a,** Quantum oscillations in field-dependent Seebeck coefficient $S_{xx}$ and resistivity $\rho_{xx}$. **b,** Landau fan diagram for the oscillations, suggesting a non-trivial Berry phase. **c,** Angle dependent oscillatory frequency and corresponding cyclotron mass for the detected orbit, both showing an inverse cosine behavior. **d-f,** The electric Hall conductivity $\sigma_{xy}$, the Nernst thermopower divided by temperature $-S_{xy}/T$ and thermal Hall conductivity divided by temperature $\kappa_{xy}/T$ at representative temperatures, showing dominant contribution of anomalous terms. **g,** The pondering function for anomalous Hall conductivity $\sigma_{xy}^A$, anomalous thermoelectric Hall conductivity $\alpha_{xy}^A$, and the anomalous thermal Hall conductivity $\kappa_{xy}^A$ at 100 K and 300 K, respectively, whose integrations over the Berry curvature amount to the three intrinsic anomalous conductivities. The Berry curvature summation $2\pi\tilde{\sigma}_{xy}(\varepsilon)$ for the Chern gapped Dirac fermion with a gap size of $\Delta$ is shown right below. **h,** Scaling of the anomalous Hall conductivity. The longitudinal conductivity $\sigma_{xx}$ for TbMn$_6$Sn$_6$ lies within the good metal region, suggesting a dominant intrinsic contribution. Inset shows a polynomial

fitting of the intrinsic Hall conductivity, amounting to 0.13 e²/h per kagome layer. **i,** The ratio $\alpha_{xy}^A/\sigma_{xy}^A$ scales with the linear function of $k_B T/E_D$, which is obtained from the Chern-gapped Dirac model. **j,** The ratio $\kappa_{xy}^A/\sigma_{xy}^A$ significantly enhances over the T-linear function expected by the Wiedemann-Franz law above 100 K, which matches the $L_0 T(1 + \eta k_B^2 T^2/E_D^2)$ behavior for the Chern-gapped Dirac model. Adapted from REF.[54].

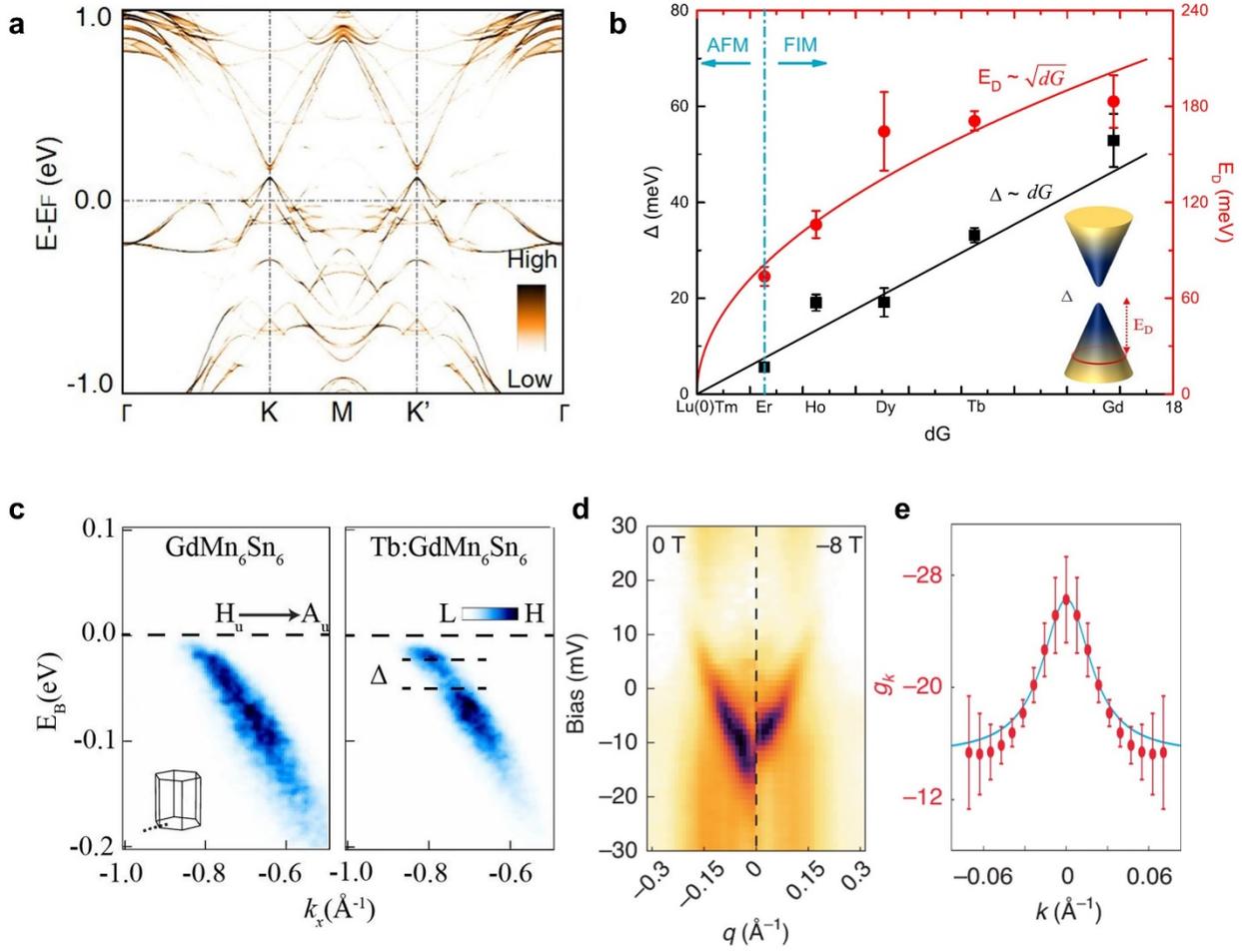

**Figure 5 Topological band engineering in RMn$_6$Sn$_6$. a,** Calculated bulk band structure of representing TbMn$_6$Sn$_6$ (momentum-resolved density of states from all orbitals along the high symmetry directions) with SOC, showing large Chern gap at $K$ and $K'$. **b,** Experimentally derived Dirac cone energy $E_D$ and gap size $\Delta$ for the whole RMn$_6$Sn$_6$ series. The systematic evolution of $\Delta$ and $E_D$ follows the de Gennes factor $dG$ and $\sqrt{dG}$, respectively. **c,** When magnetization is tuned from in-plane FiM in GdMn$_6$Sn$_6$ to out-of-plane in Tb$_{0.2}$Gd$_{0.8}$Mn$_6$Sn$_6$, the Dirac nodal line opens a SOC gap at $H_u$. **d,** Quasiparticle interference (QPI) imagings of the massive Dirac cone below Fermi level in YMn$_6$Sn$_6$ at zero field (left) and -8 T magnetic field along the $c$ direction (right). **e,** The measured $g$ factor for different electronic states in momentum space for YMn$_6$Sn$_6$. Panel **a** adapted from REF.[52]. Panel **b** adapted from REF.[53]. Panel **c** adapted from REF.[144], Panel **d** and **f** adapted from REF.[145].

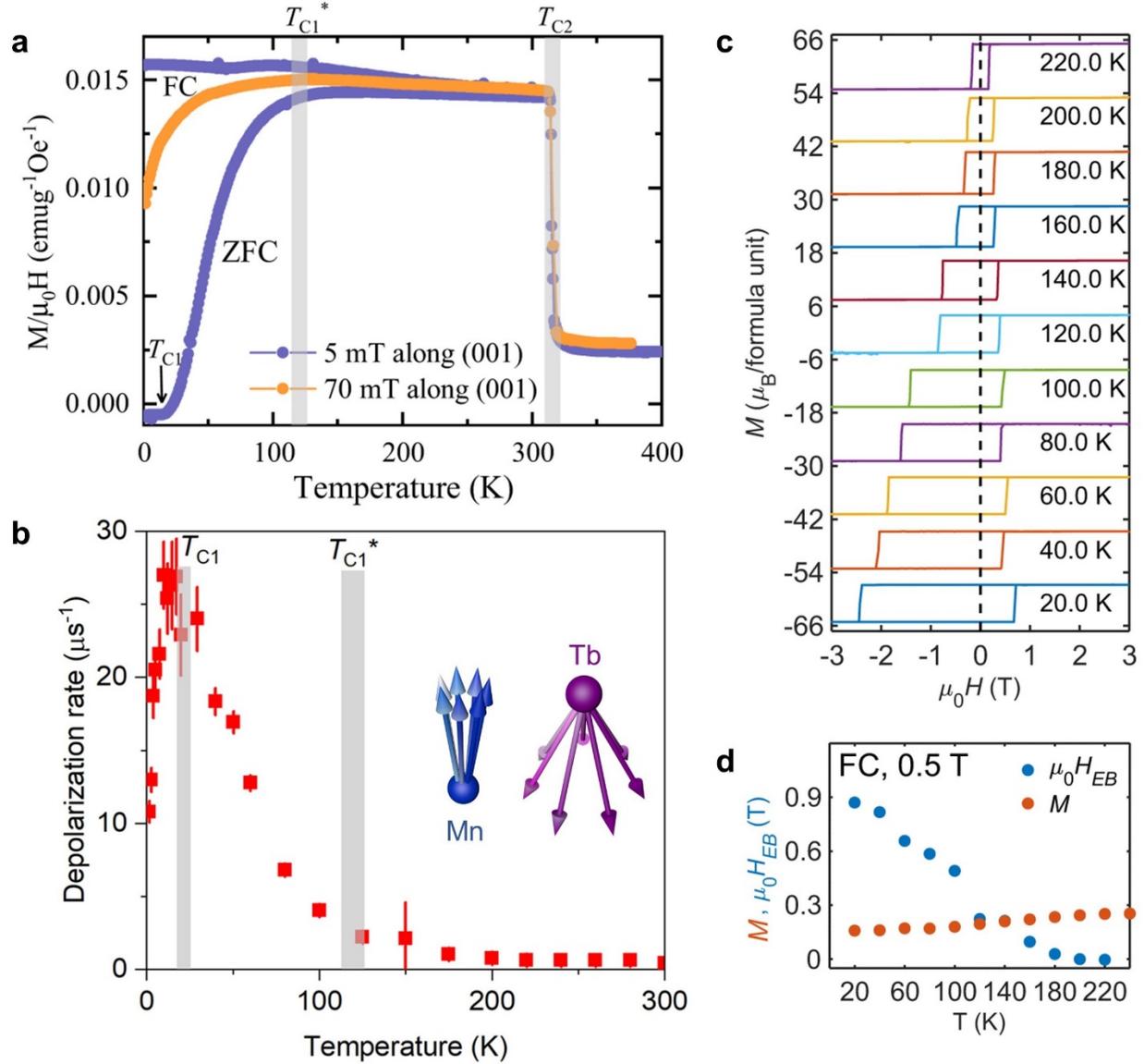

**Figure 6 Spin dynamics and exchange-bias in TbMn$_6$Sn$_6$. a,** The temperature dependence of the zero-field-cool warming and field-cool warming magnetization in TbMn$_6$Sn$_6$, showing three critical temperatures, the spin-reorientation temperature $T_{C2}$, the spin slow-down temperature $T_{C1}^*$ and the spin freezed temperature $T_{C1}$. **b,** The temperature dependence of dynamic depolarization rate of the muon-spin rotation signal. Arrows mark the magnetic transition temperature $T_{C1}$ and the temperature $T_{C1}^*$ for the onset of visible magnetic fluctuations. Inset is a cartoon illustrating spin dynamics of Mn (blue) and Tb (purple) spins. **c,** Asymmetric magnetization hysteresis loops measured at various temperatures after the sample was cooled down from 340 K to the measurement temperature with 0.5 T magnetic field prior to each measurement. **d,** Temperature dependence of the exchange biasing field $\mu_0 H_{EB}$. Panel **a** and **b** adapted from REF.[82]. Panel **c** and **d** adapted from REF.[80].

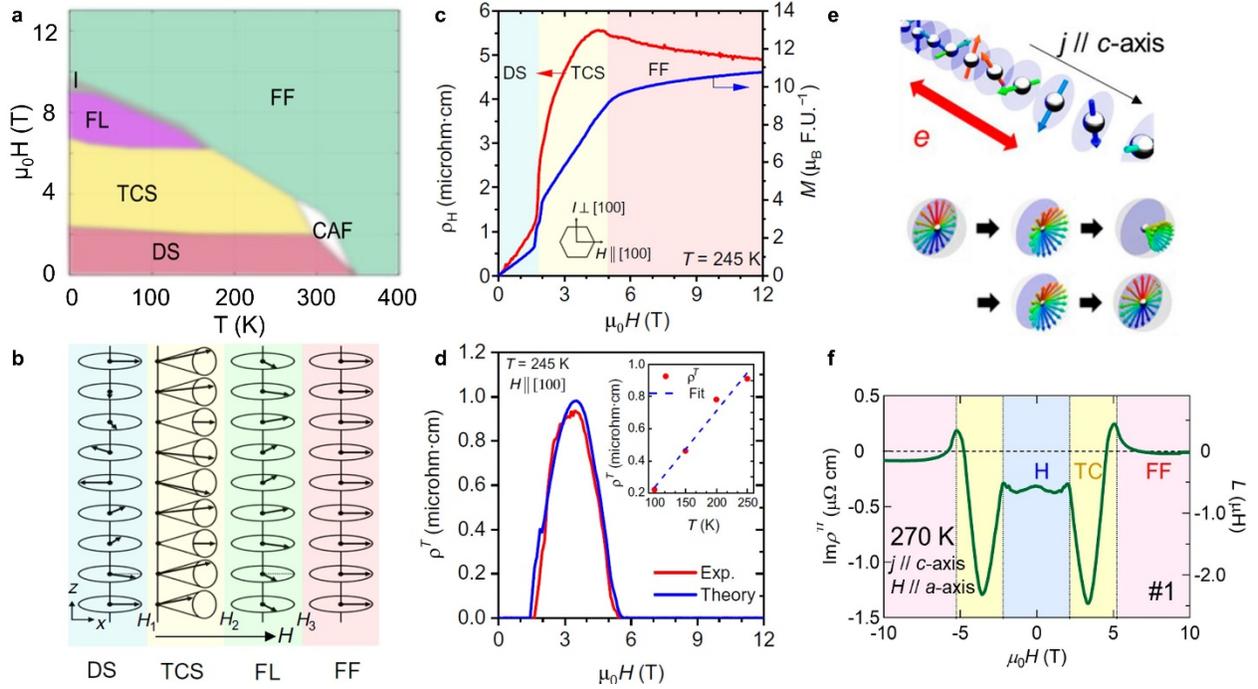

**Figure 7 Real-space Berry phase effects in YMn$_6$Sn$_6$. a,** Schematic phase diagram of YMn$_6$Sn$_6$ when the field is applied in the *ab*-plane. The various magnetic phases include the distorted spiral (DS), transverse conical spiral (TCS), fan-like (FL), and forced FM (FF) phases. CAF represents the canted AFM phase. **b,** Sketch of different magnetic structures. **c,** Hall resistivity and magnetization as a function of magnetic field applied in the *ab* plane at 245 K. **d,** The topological Hall resistivity as a function of external magnetic field at 245 K. Inset shows the temperature variation of the THE at 4 T below 250 K. **e,** Schematic illustration of emergent electromagnetic induction produced in the spin-helix state. **f,** Emergent inductance beyond room temperature and its magnetic-field dependence. Panel **a** adapted from REF.[148]. Panel **b-d** adapted from REF.[89]. Panel **e** and **f** adapted from REF.[170].

**References:**
[1] D. Xiao, M.-C. Chang, and Q. Niu, "Berry phase effects on electronic properties," Rev. Mod. Phys. 82, 1959 (2010).
[2] S. M. Girvin and K. Yang, Modern condensed matter physics (Cambridge University Press, 2019).
[3] J. E. Avron, D. Osadchy, and R. Seiler, "A topological look at the quantum Hall effect," Physics Today 56, 38 (2003).
[4] F. D. M. Haldane, "Nobel lecture: Topological quantum matter," Rev. Mod. Phys. 89, 040502 (2017).
[5] M. Z. Hasan and C. L. Kane, "Colloquium: Topological insulators," Rev. Mod. Phys. 82, 3045 (2010).
[6] X.-L. Qi and S.-C. Zhang, "Topological insulators and superconductors," Rev. Mod. Phys. 83, 1057 (2011).

[7] Y. Ando, "Topological insulator materials," J. Phys. Soc. Jpn. 82, 102001 (2013).

[8] M. Z. Hasan, S.-Y. Xu, and G. Bian, "Corrigendum: Topological insulators, topological superconductors and Weyl fermion semimetals: discoveries, perspectives and outlooks," Phys. Scr. T168, 019501 (2016).

[9] B. Bradlyn, J. Cano, Z. Wang, M. G. Vergniory, C. Felser, R. J. Cava, and B. A. Bernevig, "Beyond Dirac and Weyl fermions: Unconventional quasiparticles in conventional crystals," Science 353, aaf5037 (2016).

[10] B. Keimer and J. E. Moore, "The physics of quantum materials," Nat. Phys. 13, 1045 (2017).

[11] M. Z. Hasan, S.-Y. Xu, I. Belopolski, and S.-M. Huang, "Discovery of Weyl fermion semimetals and topological Fermi arc states," Annu. Rev. Condens. Matter Phys. 8, 289 (2017).

[12] A. Burkov, "Weyl metals," Annu. Rev. Conden. Matter Phys. 9, 359 (2018).

[13] N. P. Armitage, E. J. Mele, and A. Vishwanath, "Weyl and Dirac semimetals in three-dimensional solids," Rev. Mod. Phys. 90, 015001 (2018).

[14] S. Sachdev, "Topological order, emergent gauge fields, and Fermi surface reconstruction," Rep. Prog. Phys. 82, 014001 (2018).

[15] X.-G. Wen, "Choreographed entanglement dances: Topological states of quantum matter," Science 363, eaal3099 (2019).

[16] D. E. Kharzeev, "Topology, magnetic field, and strongly interacting matter," Annu. Rev. Nucl. Part. Sci. 65, 193 (2015).

[17] F. D. M. Haldane, "Model for a quantum Hall effect without Landau levels: Condensed-matter realization of the "parity anomaly"," Phys. Rev. Lett. 61, 2015 (1988).

[18] Y.-F. Wang, Z.-C. Gu, C.-D. Gong, and D. N. Sheng, "Fractional quantum Hall effect of hard-core bosons in topological flat bands," Phys. Rev. Lett. 107, 146803 (2011).

[19] B. Sutherland, "Localization of electronic wave functions due to local topology," Phys. Rev. B 34, 5208 (1986).

[20] C. Weeks and M. Franz, "Topological insulators on the Lieb and perovskite lattices," Phys. Rev. B 82, 085310 (2010).

[21] S. Mukherjee, A. Spracklen, D. Choudhury, N. Goldman, P. Öhberg, E. Andersson, and R. R. Thomson, "Observation of a localized flat-band state in a photonic Lieb lattice," Phys. Rev. Lett. 114, 245504 (2015).

[22] I. Syôzi, "Statistics of Kagomé Lattice," Prog. Theor. Phys. 6, 306 (1951).

[23] H.-M. Guo and M. Franz, "Topological insulator on the kagome lattice," Phys. Rev. B 80, 113102 (2009).

[24] D. Leykam, A. Andreanov, and S. Flach, "Artificial flat band systems: from lattice models to experiments," Adv. Phys.: X 3, 1473052 (2018).

[25] T. Neupert, M. M. Denner, J.-X. Yin, R. Thomale, and M. Z. Hasan, "Charge order and superconductivity in kagome materials," Nat. Phys. 18, 137 (2022).

[26] K. Ohgushi, S. Murakami, and N. Nagaosa, "Spin anisotropy and quantum Hall effect in the kagome lattice: Chiral spin state based on a ferromagnet," Phys. Rev. B 62, R6065 (2000).

[27] G. Xu, B. Lian, and S.-C. Zhang, "Intrinsic quantum anomalous Hall effect in the kagome lattice

Cs$_2$LiMn$_3$F$_{12}$," Phys. Rev. Lett. 115, 186802 (2015).

[28] C. Wu, D. Bergman, L. Balents, and S. Das Sarma, "Flat bands and Wigner crystallization in the honeycomb optical lattice," Phys. Rev. Lett. 99, 070401 (2007).

[29] M. L. Kiesel and R. Thomale, "Sublattice interference in the kagome Hubbard model," Phys. Rev. B 86, 121105 (2012).

[30] S.-L. Yu and J.-X. Li, "Chiral superconducting phase and chiral spin-density-wave phase in a Hubbard model on the kagome lattice," Phys. Rev. B 85, 144402 (2012).

[31] M. L. Kiesel, C. Platt, and R. Thomale, "Unconventional Fermi surface instabilities in the kagome Hubbard model," Phys. Rev. Lett. 110, 126405 (2013).

[32] Y.-X. Jiang, J.-X. Yin, M. M. Denner, N. Shumiya, B. R. Ortiz, G. Xu, Z. Guguchia, J. He, M. S. Hossain, X. Liu, J. Ruff, L. Kautzsch, S. S. Zhang, G. Chang, I. Belopolski, Q. Zhang, T. A. Cochran, D. Multer, M. Litskevich, Z.-J. Cheng, X. P. Yang, Z. Wang, R. Thomale, T. Neupert, S. D. Wilson, and M. Z. Hasan, "Unconventional chiral charge order in kagome superconductor KV$_3$Sb$_5$," Nat. Mater. 20, 1353 (2021).

[33] E. Tang, J.-W. Mei, and X.-G. Wen, "High-temperature fractional quantum Hall states," Phys. Rev. Lett. 106, 236802 (2011).

[34] H. Tan, Y. Liu, Z. Wang, and B. Yan, "Charge density waves and electronic properties of superconducting kagome metals," Phys. Rev. Lett. 127, 046401 (2021).

[35] M. M. Denner, R. Thomale, and T. Neupert, "Analysis of charge order in the kagome metal AV$_3$Sb$_5$ (A = K, Rb, Cs)," Phys. Rev. Lett. 127, 217601 (2021).

[36] Z. Liang, X. Hou, F. Zhang, W. Ma, P. Wu, Z. Zhang, F. Yu, J.-J. Ying, K. Jiang, L. Shan, Z. Wang, and X.-H. Chen, "Three-dimensional charge density wave and surface-dependent vortex-core states in a kagome superconductor CsV$_3$Sb$_5$," Phys. Rev. X 11, 031026 (2021).

[37] B. R. Ortiz, S. M. L. Teicher, L. Kautzsch, P. M. Sarte, N. Ratcliff, J. Harter, J. P. C. Ruff, R. Seshadri, and S. D. Wilson, "Fermi surface mapping and the nature of charge-density-wave order in the kagome superconductor CsV$_3$Sb$_5$," Phys. Rev. X 11, 041030 (2021).

[38] L. Nie, K. Sun, W. Ma, D. Song, L. Zheng, Z. Liang, P. Wu, F. Yu, J. Li, M. Shan, D. Zhao, S. Li, B. Kang, Z. Wu, Y. Zhou, K. Liu, Z. Xiang, J. Ying, Z. Wang, T. Wu, and X. Chen, "Charge-density-wave-driven electronic nematicity in a kagome superconductor," Nature 604, 59 (2022).

[39] M. Kang, S. Fang, J.-K. Kim, B. R. Ortiz, S. H. Ryu, J. Kim, J. Yoo, G. Sangiovanni, D. Di Sante, B.-G. Park, C. Jozwiak, A. Bostwick, E. Rotenberg, E. Kaxiras, S. D. Wilson, J.-H. Park, and R. Comin, "Twofold van Hove singularity and origin of charge order in topological kagome superconductor CsV$_3$Sb$_5$," Nat. Phys. 18, 301 (2022).

[40] S. Yan, D. A. Huse, and S. R. White, "Spin-liquid ground state of the S = 1/2 kagome Heisenberg antiferromagnet," Science 332, 1173 (2011).

[41] T.-H. Han, J. S. Helton, S. Chu, D. G. Nocera, J. A. Rodriguez-Rivera, C. Broholm, and Y. S. Lee, "Fractionalized excitations in the spin-liquid state of a kagome-lattice antiferromagnet," Nature 492, 406 (2012).

[42] L. Ye, M. Kang, J. Liu, F. Von Cube, C. R. Wicker, T. Suzuki, C. Jozwiak, A. Bostwick, E. Rotenberg, D. C. Bell, et al., "Massive Dirac fermions in a ferromagnetic kagome metal," Nature 555, 638 (2018).

[43] J.-X. Yin, S. S. Zhang, G. Chang, Q. Wang, S. S. Tsirkin, Z. Guguchia, B. Lian, H. Zhou, K. Jiang, I. Belopolski, et al., "Negative flat band magnetism in a spin–orbit-coupled correlated kagome magnet," Nat. Phys. 15, 443 (2019).

[44] K. Kuroda, T. Tomita, M.-T. Suzuki, C. Bareille, A. . A. Nugroho, P. Goswami, M. Ochi, M. Ikhlas, M. Nakayama, S. Akebi, R. Noguchi, R. Ishii, N. Inami, K. Ono, H. Kumigashira, A. Varykhalov, T. Muro, T. Koretsune, R. Arita, S. Shin, T. Kondo, and S. Nakatsuji, "Evidence for magnetic Weyl fermions in a correlated metal," Nat. Mater. 16, 1090 (2017).

[45] N. Morali, R. Batabyal, P. K. Nag, E. Liu, Q. Xu, Y. Sun, B. Yan, C. Felser, N. Avraham, and H. Beidenkopf, "Fermi-arc diversity on surface terminations of the magnetic Weyl semimetal $Co_3Sn_2S_2$," Science 365, 1286 (2019).

[46] S. Nakatsuji, N. Kiyohara, and T. Higo, "Large anomalous Hall effect in a non-collinear antiferromagnet at room temperature," Nature 527, 212 (2015).

[47] X. Li, L. Xu, L. Ding, J. Wang, M. Shen, X. Lu, Z. Zhu, and K. Behnia, "Anomalous Nernst and Righi-Leduc effects in $Mn_3Sn$: Berry curvature and entropy flow," Phys. Rev. Lett. 119, 056601 (2017).

[48] A. K. Nayak, J. E. Fischer, Y. Sun, B. Yan, J. Karel, A. C. Komarek, C. Shekhar, N. Kumar, W. Schnelle, J. Kbler, C. Felser, and S. S. P. Parkin, "Large anomalous Hall effect driven by a nonvanishing Berry curvature in the noncolinear antiferromagnet $Mn_3Ge$," Sci. Adv. 2, e1501870 (2016).

[49] E. Liu, Y. Sun, N. Kumar, L. Muechler, A. Sun, L. Jiao, S.-Y. Yang, D. Liu, A. Liang, Q. Xu, et al., "Giant anomalous Hall effect in a ferromagnetic kagome-lattice semimetal," Nat. Phys. 14, 1125 (2018).

[50] Q. Wang, Y. Xu, R. Lou, Z. Liu, M. Li, Y. Huang, D. Shen, H. Weng, S. Wang, and H. Lei, "Large intrinsic anomalous Hall effect in half-metallic ferromagnet $Co_3Sn_2S_2$ with magnetic Weyl fermions," Nat. Commun. 9, 3681 (2018).

[51] T. Kida, L. Fenner, A. Dee, I. Terasaki, M. Hagiwara, and A. Wills, "The giant anomalous Hall effect in the ferromagnet $Fe_3Sn_2$ – a frustrated kagome metal," J. Phys. Condens. Matter 23, 112205 (2011).

[52] J.-X. Yin, W. Ma, T. A. Cochran, X. Xu, S. S. Zhang, H.-J. Tien, N. Shumiya, G. Cheng, K. Jiang, B. Lian, et al., "Quantum-limit Chern topological magnetism in $TbMn_6Sn_6$," Nature 583, 533 (2020).

[53] W. Ma, X. Xu, J.-X. Yin, H. Yang, H. Zhou, Z.-J. Cheng, Y. Huang, Z. Qu, F. Wang, M. Z. Hasan, and S. Jia, "Rare earth engineering in $RMn_6Sn_6$ (R = Gd - Tm, Lu) topological kagome magnets," Phys. Rev. Lett. 126, 246602 (2021).

[54] X. Xu, J.-X. Yin, W. Ma, H.-J. Tien, X.-B. Qiang, P. V. S. Reddy, H. Zhou, J. Shen, H.-Z. Lu, T.-R. Chang, Z. Qu, and S. Jia, "Topological charge-entropy scaling in kagome Chern magnet $TbMn_6Sn_6$," Nat. Commun. 13, 1197 (2022).

[55] S.-Y. Yang, Y. Wang, B. R. Ortiz, D. Liu, J. Gayles, E. Derunova, R. Gonzalez-Hernandez, L. ˇSmejkal, Y. Chen, S. S. P. Parkin, S. D. Wilson, E. S. Toberer, T. McQueen, and M. N. Ali, "Giant, unconventional anomalous Hall effect in the metallic frustrated magnet candidate, $KV_3Sb_5$," Sci. Adv. 6, eabb6003 (2020).

[56] F. H. Yu, T. Wu, Z. Y. Wang, B. Lei, W. Z. Zhuo, J. J. Ying, and X. H. Chen, "Concurrence of anomalous

Hall effect and charge density wave in a superconducting topological kagome metal," Phys. Rev. B 104, L041103 (2021).

[57] M. P. Shores, E. A. Nytko, B. M. Bartlett, and D. G. Nocera, "A structurally perfect S = 1/2 kagomé antiferromagnet," J. Am. Chem. Soc. 127, 13462 (2005).

[58] J. S. Helton, K. Matan, M. P. Shores, E. A. Nytko, B. M. Bartlett, Y. Yoshida, Y. Takano, A. Suslov, Y. Qiu, J.-H. Chung, D. G. Nocera, and Y. S. Lee, "Spin dynamics of the spin-1/2 kagome lattice antiferromagnet $ZnCu_3(OH)_6Cl_2$," Phys. Rev. Lett. 98, 107204 (2007).

[59] U. Englich, C. Frommen, and W. Massa, "Jahn-Teller ordering in Kagomé-type layers of compounds $A_2AMn^{III}_3F_{12}$ (A=Rb, Cs; A=Li, Na, K)," J. Alloy Compd. 246, 155 (1997).

[60] N. Kiyohara, T. Tomita, and S. Nakatsuji, "Giant anomalous Hall effect in the chiral antiferromagnet $Mn_3Ge$," Phys. Rev. Applied 5, 064009 (2016).

[61] Z. Lin, J.-H. Choi, Q. Zhang, W. Qin, S. Yi, P. Wang, L. Li, Y. Wang, H. Zhang, Z. Sun, L. Wei, S. Zhang, T. Guo, Q. Lu, J.-H. Cho, C. Zeng, and Z. Zhang, "Flatbands and emergent ferromagnetic ordering in $Fe_3Sn_2$ kagome lattices," Phys. Rev. Lett. 121, 096401 (2018).

[62] M. Kang, L. Ye, S. Fang, J.-S. You, A. Levitan, M. Han, J. I. Facio, C. Jozwiak, A. Bostwick, E. Rotenberg, M. K. Chan, R. D. McDonald, D. Graf, K. Kaznatcheev, E. Vescovo, D. C. Bell, E. Kaxiras, J. van den Brink, M. Richter, M. Prasad Ghimire, J. G. Checkelsky, and R. Comin, "Dirac fermions and flat bands in the ideal kagome metal FeSn," Nat. Mater. 19, 163 (2020).

[63] Z. Liu, M. Li, Q. Wang, G. Wang, C. Wen, K. Jiang, X. Lu, S. Yan, Y. Huang, D. Shen, J.-X. Yin, Z. Wang, Z. Yin, H. Lei, and S. Wang, "Orbital-selective Dirac fermions and extremely flat bands in frustrated kagome- lattice metal CoSn," Nat. Commun. 11, 4002 (2020).

[64] X. Teng, L. Chen, F. Ye, E. Rosenberg, Z. Liu, J.- X. Yin, Y.-X. Jiang, J. S. Oh, M. Z. Hasan, K. J. Neubauer, et al., "Discovery of charge density wave in a correlated kagome lattice antiferromagnet," arXiv preprint arXiv:2203.11467 (2022).

[65] S. Shao, J.-X. Yin, I. Belopolski, J.-Y. You, T. Hou, H. Chen, Y.-X. Jiang, M. S. Hossain, M. Yahyavi, C.- H. Hsu, et al., "Charge density wave interaction in a kagome-honeycomb antiferromagnet," arXiv preprint arXiv:2206.12033 (2022).

[66] Y. Ishii, H. Harima, Y. Okamoto, J.-i. Yamaura, and Z. Hiroi, "$YCr_6Ge_6$ as a candidate compound for a kagome metal," J. Phys. Soc. Jpn. 82, 023705 (2013).

[67] S. Peng, Y. Han, G. Pokharel, J. Shen, Z. Li, M. Hashimoto, D. Lu, B. R. Ortiz, Y. Luo, H. Li, M. Guo, B. Wang, S. Cui, Z. Sun, Z. Qiao, S. D. Wilson, and J. He, "Realizing kagome band structure in two-dimensional kagome surface states of $RV_6Sn_6$ (R = Gd, Ho)," Phys. Rev. Lett. 127, 266401 (2021).

[68] Y. Hu, X. Wu, Y. Yang, S. Gao, N. C. Plumb, A. P. Schnyder, W. Xie, J. Ma, and M. Shi, "Tunable topological Dirac surface states and van Hove singularities in kagome metal $GdV_6Sn_6$," Sci. Adv. 8, eadd2024 (2022).

[69] H. W. S. Arachchige, W. R. Meier, M. Marshall, T. Matsuoka, R. Xue, M. A. McGuire, R. P. Hermann, H. Cao, and D. Mandrus, "Charge density wave in kagome lattice intermetallic $ScV_6Sn_6$," Phys. Rev. Lett. 129, 216402 (2022).

[70] X. Zhang, J. Hou, W. Xia, Z. Xu, P. Yang, A. Wang, Z. Liu, J. Shen, H. Zhang, X. Dong, Y. Uwatoko, J. Sun, B. Wang, Y. Guo, and J. Cheng, "Destabilization of the charge density wave and the absence of superconductivity in $ScV_6Sn_6$ under high pressures up to 11 GPa," Materials 15 (2022), 10.3390/ma15207372.

[71] T. Hu, S. Xu, L. Yue, Q. Wu, Q. Liu, S. Zhang, R. Li, X. Zhou, J. Yuan, D. Wu, T. Dong, and N. Wang, "Optical studies of structural phase transition in the vanadium-based kagome metal $ScV_6Sn_6$," arXiv preprint arXiv:2211.03412 (2022).

[72] Q. Yin, Z. Tu, C. Gong, Y. Fu, S. Yan, and H. Lei, "Superconductivity and normal-state properties of kagome metal $RV_3Sb_5$ single crystals," Chinese Phys. Lett. 38, 037403 (2021).

[73] S. Cho, H. Ma, W. Xia, Y. Yang, Z. Liu, Z. Huang, Z. Jiang, X. Lu, J. Liu, Z. Liu, J. Li, J. Wang, Y. Liu, J. Jia, Y. Guo, J. Liu, and D. Shen, "Emergence of new van Hove singularities in the charge density wave state of a topological kagome metal $RV_3Sb_5$," Phys. Rev. Lett. 127, 236401 (2021).

[74] H. Zhao, H. Li, B. R. Ortiz, S. M. L. Teicher, T. Park, M. Ye, Z. Wang, L. Balents, S. D. Wilson, and I. Zeljkovic, "Cascade of correlated electron states in the kagome superconductor $CsV_3Sb_5$," Nature 599, 216 (2021).

[75] C. Mielke, Y. Qin, J.-X. Yin, H. Nakamura, D. Das, K. Guo, R. Khasanov, J. Chang, Z. Q. Wang, S. Jia, S. Nakatsuji, A. Amato, H. Luetkens, G. Xu, M. Z. Hasan, and Z. Guguchia, "Nodeless kagome superconductivity in $LaRu_3Si_2$," Phys. Rev. Materials 5, 034803 (2021).

[76] S. S. Banerjee, N. G. Patil, S. Saha, S. Ramakrishnan, A. K. Grover, S. Bhattacharya, G. Ravikumar, P. K. Mishra, T. V. Chandrasekhar Rao, V. C. Sahni, M. J. Higgins, E. Yamamoto, Y. Haga, M. Hedo, Y. Inada, and Y. Onuki, "Anomalous peak effect in $CeRu_2$: Fracturing of a flux line lattice," Phys. Rev. B 58, 995 (1998).

[77] L. Gao, S. Shen, Q. Wang, W. Shi, Y. Zhao, C. Li, W. Cao, C. Pei, J.-Y. Ge, G. Li, J. Li, Y. Chen, S. Yan, and Y. Qi, "Anomalous Hall effect in ferrimagnetic metal $RMn_6Sn_6$ (R = Tb, Dy, Ho) with clean Mn kagome lattice," Appl. Phys. Lett. 119, 092405 (2021).

[78] G. Dhakal, F. Cheenicode Kabeer, A. K. Pathak, F. Kabir, N. Poudel, R. Filippone, J. Casey, A. Pradhan Sakhya, S. Regmi, C. Sims, K. Dimitri, P. Manfrinetti, K. Gofryk, P. M. Oppeneer, and M. Neupane, "Anisotropically large anomalous and topological Hall effect in a kagome magnet," Phys. Rev. B 104, L161115 (2021).

[79] T. Asaba, S. M. Thomas, M. Curtis, J. D. Thompson, E. D. Bauer, and F. Ronning, "Anomalous Hall effect in the kagome ferrimagnet $GdMn_6Sn_6$," Phys. Rev. B 101, 174415 (2020).

[80] H. Zhang, J. Koo, C. Xu, M. Sretenovic, B. Yan, and X. Ke, "Exchange-biased topological transverse thermoelectric effects in a kagome ferrimagnet," Nat. Commun. 13, 1091 (2022).

[81] S. X. M. Riberolles, T. J. Slade, D. L. Abernathy, G. E. Granroth, B. Li, Y. Lee, P. C. Canfield, B. G. Ueland, L. Ke, and R. J. McQueeney, "Low-temperature competing magnetic energy scales in the topological ferrimagnet $TbMn_6Sn_6$," Phys. Rev. X 12, 021043 (2022).

[82] C. Mielke III, W. L. Ma, V. Pomjakushin, O. Zaharko, S. Sturniolo, X. Liu, V. Ukleev, J. S. White, J.-X. Yin, S. S. Tsirkin, C. B. Larsen, T. A. Cochran, M. Medarde, V. Porée, D. Das, R. Gupta, C. N. Wang, J.

Chang, Z. Q. Wang, R. Khasanov, T. Neupert, A. Amato, L. Liborio, S. Jia, M. Z. Hasan, H. Luetkens, and Z. Guguchia, "Low-temperature magnetic crossover in the topological kagome magnet TbMn$_6$Sn$_6$," Commun. Phys. 5, 107 (2022).

[83] D. C. Jones, S. Das, H. Bhandari, X. Liu, P. Siegfried, M. P. Ghimire, S. S. Tsirkin, I. Mazin, and N. J. Ghimire, "Origin of spin reorientation and intrinsic anomalous Hall effect in the kagome ferrimagnet TbMn$_6$Sn$_6$," arXiv preprint arXiv:2203.17246 (2022).

[84] Y. Lee, R. Skomski, X. Wang, P. Orth, A. Pathak, B. Harmon, R. McQueeney, and L. Ke, "Interplay between magnetism and band topology in kagome magnets RMn$_6$Sn$_6$," arXiv preprint arXiv:2201.11265 (2022).

[85] X. Gu, C. Chen, W. S. Wei, L. L. Gao, J. Y. Liu, X. Du, D. Pei, J. S. Zhou, R. Z. Xu, Z. X. Yin, W. X. Zhao, Y. D. Li, C. Jozwiak, A. Bostwick, E. Rotenberg, D. Backes, L. S. I. Veiga, S. Dhesi, T. Hesjedal, G. van der Laan, H. F. Du, W. J. Jiang, Y. P. Qi, G. Li, W. J. Shi, Z. K. Liu, Y. L. Chen, and L. X. Yang, "Robust kagome electronic structure in the topological quantum magnets XMn$_6$Sn$_6$ (X = Dy, Tb, Gd, Y)," Phys. Rev. B 105, 155108 (2022).

[86] R. Li, T. Zhang, W. Ma, S. Xu, Q. Wu, L. Yue, S. Zhang, Q. Liu, Z. Wang, T. Hu, et al., "Flat optical conductivity in topological kagome magnet TbMn$_6$Sn$_6$," arXiv preprint arXiv:2207.09308 (2022).

[87] M. Wenzel, O. Iakutkina, Q. Yin, H. Lei, M. Dressel, and E. Uykur, "Effect of magnetism and phonons on localized carriers in ferrimagnetic kagome metals GdMn$_6$Sn$_6$ and TbMn$_6$Sn$_6$," arXiv preprint arXiv:2208.00756 (2022).

[88] H. Zhang, X. Feng, T. Heitmann, A. I. Kolesnikov, M. B. Stone, Y.-M. Lu, and X. Ke, "Topological magnon bands in a room-temperature kagome magnet," Phys. Rev. B 101, 100405 (2020).

[89] N. J. Ghimire, R. L. Dally, L. Poudel, D. Jones, D. Michel, N. T. Magar, M. Bleuel, M. A. McGuire, J. Jiang, J. Mitchell, et al., "Competing magnetic phases and fluctuation-driven scalar spin chirality in the kagome metal YMn$_6$Sn$_6$," Sci. Adv. 6, eabe2680 (2020).

[90] M. Li, Q. Wang, G. Wang, Z. Yuan, W. Song, R. Lou, Z. Liu, Y. Huang, Z. Liu, H. Lei, Z. Yin, and S. Wang, "Dirac cone, flat band and saddle point in kagome magnet YMn$_6$Sn$_6$," Nat. Commun. 12, 3129 (2021).

[91] Q. Wang, K. J. Neubauer, C. Duan, Q. Yin, S. Fujitsu, H. Hosono, F. Ye, R. Zhang, S. Chi, K. Krycka, H. Lei, and P. Dai, "Field-induced topological Hall effect and double-fan spin structure with a *c*-axis component in the metallic kagome antiferromagnetic compound YMn$_6$Sn$_6$," Phys. Rev. B 103, 014416 (2021).

[92] R. L. Dally, J. W. Lynn, N. J. Ghimire, D. Michel, P. Siegfried, and I. I. Mazin, "Chiral properties of the zero-field spiral state and field-induced magnetic phases of the itinerant kagome metal YMn$_6$Sn$_6$," Phys. Rev. B 103, 094413 (2021).

[93] H. Zeng, G. Yu, X. Luo, C. Chen, C. Fang, S. Ma, Z. Mo, J. Shen, M. Yuan, and Z. Zhong, "Large anomalous Hall effect in kagomé ferrimagnetic HoMn$_6$Sn$_6$ single crystal," J. Alloy. Compd. 899, 163356 (2022).

[94] B. Wang, E. Yi, L. Li, J. Qin, B.-F. Hu, B. Shen, and M. Wang, "Magneto-transport properties of

kagome magnet TmMn$_6$Sn$_6$," arXiv preprint arXiv:2205.07806 (2022).

[95] G. Venturini, "Filling the CoSn host-cell: the HfFe$_6$Ge$_6$-type and the related structures," Z. Krist.-Cryst. Mater. 221, 511 (2006).

[96] D. C. Fredrickson, S. Lidin, G. Venturini, B. Malaman, and J. Christensen, "Origins of superstructure ordering and incommensurability in stuffed CoSn-type phases," J. Am. Chem. Soc. 130, 8195 (2008).

[97] N. V. Baranov, E. G. Gerasimov, and N. V. Mushnikov, "Magnetism of compounds with a layered crystal structure," Phys. Met. Metallogr. 112, 711 (2011).

[98] D. Clatterbuck and K. Gschneidner, "Magnetic properties of RMn$_6$Sn$_6$ (R = Tb, Ho, Er, Tm, Lu) single crystals," J. Magn. Magn. Mater. 207, 78 (1999).

[99] P. C. Canfield and Z. Fisk, "Growth of single crystals from metallic fluxes," Phil. Mag. B 65, 1117 (1992).

[100] M. A. Avila, T. Takabatake, Y. Takahashi, S. L. Bud'ko, and P. C. Canfield, "Direct observation of Fe spin reorientation in single-crystalline YbFe$_6$Ge$_6$," J. Phys. Condens. Matter 17, 6969 (2005).

[101] L. Zhang, Unusual Magnetic Behavior of Some Rare-earth and Manganese Compounds (Universiteit van Amsterdam [Host], 2005).

[102] G. Venturini, B. E. Idrissi, and B. Malaman, "Magnetic properties of RMn$_6$Sn$_6$ (R = Sc, Y, Gd-Tm, Lu) compounds with HfFe$_6$Ge$_6$ type structure," J. Magn. Magn. Mater. 94, 35 (1991).

[103] Y. Wang, D. Wiards, D. Ryan, and J. Cadogan, "Structural and magnetic properties of RFe$_6$Ge$_6$ (R = Y, Gd, Tb, Er)," IEEE T. Magn. 30, 4951 (1994).

[104] W. Buchholz and H.-U. Schuster, "Intermetallische phasen mit b35-Überstruktur und verwandtschafts-beziehung zu LiFe$_6$Ge$_6$," Z. Anorg. Allg. Chem. 482, 40 (1981).

[105] J.-X. Yin, S. H. Pan, and M. Z. Hasan, "Probing topological quantum matter with scanning tunnelling microscopy," Nat. Rev. Phys. 3, 249 (2021).

[106] G. Venturini, R. Welter, and B. Malaman, "Crystallographic data and magnetic properties of RT$_6$Ge$_6$ compounds (R = Sc, Y, Nd, Sm, Gd-Lu; T=Mn, Fe)," J. Alloy. Compd. 185, 99 (1992).

[107] J. Brabers, V. Duijn, F. de Boer, and K. Buschow, "Magnetic properties of rare-earth manganese compounds of the RMn$_6$Ge$_6$ type," J. Alloy. Compd. 198, 127 (1993).

[108] J. Cadogan and D. Ryan, "Independent magnetic ordering of the rare-earth (R) and Fe sublattices in the RFe$_6$Ge$_6$ and RFe$_6$Sn$_6$ series," J. Alloy. Compd. 326, 166 (2001), proceedings of the International Conference on Magnetic Materials (ICMM).

[109] G. Venturini, R. Welter, B. Malaman, and E. Ressouche, "Magnetic structure of YMn$_6$Ge$_6$ and room temperature magnetic structure of LuMn$_6$Sn$_6$ obtained from neutron diffraction study," J. Alloy. Compd. 200, 51 (1993).

[110] G. Venturini, D. Fruchart, and B. Malaman, "Incommensurate magnetic structures of RMn$_6$Sn$_6$ (R = Sc, Y, Lu) compounds from neutron diffraction study," J. Alloy. Compd. 236, 102 (1996).

[111] E. Rosenfeld and N. Mushnikov, "Double-flat-spiral magnetic structures: Theory and application to the RMn$_6$X$_6$ compounds," Physica B 403, 1898 (2008).

[112] F. Weitzer, A. Leithe-Jasper, K. Hiebl, P. Rogl, Q. Qi, and J. M. D. Coey, "Structural chemistry,

magnetism and $^{119}$Sn Mössbauer spectroscopy of ternary compounds REMn$_6$Sn$_6$ (RE = Pr, Nd, Sm)," J. Appl. Phys. 73, 8447 (1993).

[113] B. Malaman, G. Venturini, B. Chafik El Idrissi, and E. Ressouche, "Magnetic properties of NdMn$_6$Sn$_6$ and SmMn$_6$Sn$_6$ compounds from susceptibility measurements and neutron diffraction study," J. Alloy. Compd. 252, 41 (1997).

[114] W. Ma, X. Xu, Z. Wang, H. Zhou, M. Marshall, Z. Qu, W. Xie, and S. Jia, "Anomalous Hall effect in the distorted kagome magnets (Nd,Sm)Mn$_6$Sn$_6$," Phys. Rev. B 103, 235109 (2021).

[115] S. Sinnema, R. Radwanski, J. Franse, D. de Mooij, and K. Buschow, "Magnetic properties of ternary rare-earth compounds of the type R$_2$Fe$_{14}$B," J. Magn. Magn. Mater. 44, 333 (1984).

[116] M. Brooks, L. Nordström, and B. Johansson, "Rare-earth transition-metal intermetallics," Physica B: Condens. Matter 172, 95 (1991).

[117] B. Malaman, G. Venturini, R. Welter, J. Sanchez, P. Vulliet, and E. Ressouche, "Magnetic properties of RMn$_6$Sn$_6$ (R = Gd-Er) compounds from neutron diffraction and Mössbauer measurements," J. Magn. Magn. Mater. 202, 519 (1999).

[118] D. Chen, C. Le, C. Fu, H. Lin, W. Schnelle, Y. Sun, and C. Felser, "Large anomalous Hall effect in the kagome ferromagnet LiMn$_6$Sn$_6$," Phys. Rev. B 103, 144410 (2021).

[119] T. Mazet, G. Venturini, R. Welter, and B. Malaman, "A magnetic study of Mg$_{1-x}$Ca$_x$Mn$_6$Sn$_6$ compounds (0.0 ≤ x ≤ 0.7).: First example of ferromagnetic HfFe$_6$Ge$_6$-type structure compounds," J. Alloy. Compd. 264, 71 (1998).

[120] T. Mazet, R. Welter, and B. Malaman, "A study of the new ferromagnetic YbMn$_6$Sn$_6$ compound by magnetization and neutron diffraction measurements," J. Magn. Magn. Mater. 204, 11 (1999).

[121] I. Giaever, "Electron tunneling and superconductivity," Science 183, 1253 (1974).

[122] J. Tersoff and D. R. Hamann, "Theory of the scanning tunneling microscope," Phys. Rev. B 31, 805 (1985).

[123] E. Merzbacher, "The early history of quantum tunneling," Phys. Today 55, 44 (2002).

[124] J.-X. Yin, S. S. Zhang, H. Li, K. Jiang, G. Chang, B. Zhang, B. Lian, C. Xiang, I. Belopolski, H. Zheng, et al., "Giant and anisotropic many-body spin–orbit tunability in a strongly correlated kagome magnet," Nature 562, 91 (2018).

[125] L. Jiao, Q. Xu, Y. Cheon, Y. Sun, C. Felser, E. Liu, and S. Wirth, "Signatures for half-metallicity and nontrivial surface states in the kagome lattice Weyl semimetal Co$_3$Sn$_2$S$_2$," Phys. Rev. B 99, 245158 (2019).

[126] H.-H. Yang, C.-C. Lee, Y. Yoshida, M. Ikhlas, T. Tomita, A. Nugroho, T. Ozaki, S. Nakatsuji, and Y. Hasegawa, "Scanning tunneling microscopy on cleaved Mn$_3$Sn (0001) surface," Sci. Rep. 9, 9677 (2019).

[127] R. Fletcher, "On the amplitude of the quantum oscillations in the thermopower of metals," J. Low Temp. Phys. 43, 363 (1981).

[128] D. Shoenberg, Magnetic Oscillations in Metals (Cambridge university press, 2009).

[129] L. Ye, M. K. Chan, R. D. McDonald, D. Graf, M. Kang, J. Liu, T. Suzuki, R. Comin, L. Fu, and J. G. Checkelsky, "de Haas-van Alphen effect of correlated Dirac states in kagome metal Fe$_3$Sn$_2$," Nat. Commun. 10, 4870 (2019).

[130] C. M. Wang, H.-Z. Lu, and S.-Q. Shen, "Anomalous phase shift of quantum oscillations in 3D topological semimetals," Phys. Rev. Lett. 117, 077201 (2016).

[131] H. Murakawa, M. Bahramy, M. Tokunaga, Y. Kohama, C. Bell, Y. Kaneko, N. Nagaosa, H. Hwang, and Y. Tokura, "Detection of Berry's phase in a bulk Rashba semiconductor," Science 342, 1490 (2013).

[132] O. Vafek, A. Melikyan, and Z. Teˇsanoviˊc, "Quasiparticle Hall transport of d-wave superconductors in the vortex state," Phys. Rev. B 64, 224508 (2001).

[133] K. Behnia, Fundamentals of thermoelectricity (OUP Oxford, 2015).

[134] L. Xu, X. Li, X. Lu, C. Collignon, H. Fu, J. Koo, B. Fauqué, B. Yan, Z. Zhu, and K. Behnia, "Finite-temperature violation of the anomalous transverse Wiedemann-Franz law," Sci. Adv. 6, eaaz3522 (2020).

[135] A. Sakai, Y. P. Mizuta, A. A. Nugroho, R. Sihombing, T. Koretsune, M.-T. Suzuki, N. Takemori, R. Ishii, D. Nishio-Hamane, R. Arita, P. Goswami, and S. Nakatsuji, "Giant anomalous Nernst effect and quantum-critical scaling in a ferromagnetic semimetal," Nat. Phys. 14, 1119 (2018).

[136] L. Xu, X. Li, L. Ding, T. Chen, A. Sakai, B. Fauqué, S. Nakatsuji, Z. Zhu, and K. Behnia, "Anomalous transverse response of $Co_2MnGa$ and universality of the room-temperature $\alpha^A_{ij}/\sigma^A_{ij}$ ratio across topological magnets," Phys. Rev. B 101, 180404 (2020).

[137] H. Yang, W. You, J. Wang, J. Huang, C. Xi, X. Xu, C. Cao, M. Tian, Z.-A. Xu, J. Dai, and Y. Li, "Giant anomalous Nernst effect in the magnetic Weyl semimetal $Co_3Sn_2S_2$," Phys. Rev. Materials 4, 024202 (2020).

[138] L. Ding, J. Koo, L. Xu, X. Li, X. Lu, L. Zhao, Q. Wang, Q. Yin, H. Lei, B. Yan, Z. Zhu, and K. Behnia, "Intrinsic anomalous Nernst effect amplified by disorder in a half-metallic semimetal," Phys. Rev. X 9, 041061 (2019).

[139] N. A. Sinitsyn, A. H. MacDonald, T. Jungwirth, V. K. Dugaev, and J. Sinova, "Anomalous Hall effect in a two-dimensional Dirac band: The link between the Kubo-Streda formula and the semiclassical Boltzmann equation approach," Phys. Rev. B 75, 045315 (2007).

[140] N. Nagaosa, J. Sinova, S. Onoda, A. H. MacDonald, and N. P. Ong, "Anomalous Hall effect," Rev. Mod. Phys. 82, 1539 (2010).

[141] N. F. Mott, H. Jones, H. Jones, and H. Jones, The Theory of the Properties of Metals and Alloys (Courier Dover Publications, 1958).

[142] G. Wiedemann and R. Franz, "Relative conductivity of solids," Ann. Phys. Chemie 89, 497 (1853).

[143] Z. Wang, R. Boyack, and K. Levin, "Heat-bath approach to anomalous thermal transport: Effects of inelastic scattering," Phys. Rev. B 105, 134302 (2022).

[144] Z.-J. Cheng, I. Belopolski, H.-J. Tien, T. A. Cochran, X. P. Yang, W. Ma, J.-X. Yin, D. Chen, J. Zhang, C. Jozwiak, A. Bostwick, E. Rotenberg, G. Cheng, M. S. Hossain, Q. Zhang, M. Litskevich, Y.-X. Jiang, N. Yao, N. B. M. Schroeter, V. N. Strocov, B. Lian, C. Felser, G. Chang, S. Jia, T.-R. Chang, and M. Z. Hasan, "Visualization of tunable Weyl line in A-A stacking kagome magnets," Adv. Mater. n/a, 2205927.

[145] H. Li, H. Zhao, K. Jiang, Q. Wang, Q. Yin, N.-N. Zhao, K. Liu, Z. Wang, H. Lei, and I. Zeljkovic, "Manipulation of Dirac band curvature and momentum-dependent $g$ factor in a kagome magnet," Nat. Phys. 18, 644 (2022).

[146] P. Terentev and N. Mushnikov, "Magnetic anisotropy of the TbMn$_6$Sn$_6$ and GdMn$_6$Sn$_6$ compounds," Phys. Met. Metallogr. 100, 571 (2005).

[147] Z. Liu, N. Zhao, M. Li, Q. Yin, Q. Wang, Z. Liu, D. Shen, Y. Huang, H. Lei, K. Liu, and S. Wang, "Electronic correlation effects in the kagome magnet GdMn$_6$Sn$_6$," Phys. Rev. B 104, 115122 (2021).

[148] P. E. Siegfried, H. Bhandari, D. C. Jones, M. P. Ghimire, R. L. Dally, L. Poudel, M. Bleuel, J. W. Lynn, I. I. Mazin, and N. J. Ghimire, "Magnetization-driven Lifshitz transition and charge-spin coupling in the kagome metal YMn$_6$Sn$_6$," Commun. Phys. 5, 58 (2022).

[149] Q. Wang, Y. Xu, R. Lou, Z. Liu, M. Li, Y. Huang, D. Shen, H. Weng, S. Wang, and H. Lei, "Large intrinsic anomalous Hall effect in half-metallic ferromagnet Co$_3$Sn$_2$S$_2$ with magnetic Weyl fermions," Nat. Commun. 9, 1 (2018).

[150] M. A. Kassem, Y. Tabata, T. Waki, and H. Nakamura, "Low-field anomalous magnetic phase in the kagome-lattice shandite Co$_3$Sn$_2$S$_2$," Phys. Rev. B 96, 014429 (2017).

[151] Z. Guguchia, J. A. T. Verezhak, D. J. Gawryluk, S. S. Tsirkin, J.-X. Yin, I. Belopolski, H. Zhou, G. Simutis, S.-S. Zhang, T. A. Cochran, G. Chang, E. Pomjakushina, L. Keller, Z. Skrzeczkowska, Q. Wang, H. C. Lei, R. Khasanov, A. Amato, S. Jia, T. Neupert, H. Luetkens, and M. Z. Hasan, "Tunable anomalous Hall conductivity through volume-wise magnetic competition in a topological kagome magnet," Nat. Commun. 11, 559 (2020).

[152] H. Wu, P. Sun, D. Hsieh, H. Chen, D. C. Kakarla, L. Deng, C. Chu, and H. Yang, "Observation of skyrmion-like magnetism in magnetic Weyl semimetal Co$_3$Sn$_2$S$_2$," Mater. Today Phys. 12, 100189 (2020).

[153] C. Lee, P. Vir, K. Manna, C. Shekhar, J. E. Moore, M. A. Kastner, C. Felser, and J. Orenstein, "Observation of a phase transition within the domain walls of ferromagnetic Co$_3$Sn$_2$S$_2$," Nat. Commun. 13, 3000 (2022).

[154] Q. Zhang, Y. Zhang, M. Matsuda, V. O. Garlea, J. Yan, M. A. McGuire, D. A. Tennant, and S. Okamoto, "Hidden local symmetry breaking in a kagome-lattice magnetic Weyl semimetal," J. Am. Chem. Soc. 144, 14339 (2022), pMID: 35901238.

[155] Z. Shen, X. Zhu, R. Ullah, P. Klavins, V. Taufour, et al., "Magnetic domain walls depinning and the magnetization anomaly within the ferromagnetic phase of the Weyl semimetal Co$_3$Sn$_2$S$_2$," arXiv preprint arXiv:2205.06420 (2022).

[156] E. Lachman, R. A. Murphy, N. Maksimovic, R. Kealhofer, S. Haley, R. D. McDonald, J. R. Long, and J. G. Analytis, "Exchange biased anomalous Hall effect driven by frustration in a magnetic kagome lattice," Nat. Commun. 11, 560 (2020).

[157] A. Noah, F. Toric, T. D. Feld, G. Zissman, A. Gutfreund, D. Tsruya, T. R. Devidas, H. Alpern, A. Vakahi, H. Steinberg, M. E. Huber, J. G. Analytis, S. Gazit, E. Lachman, and Y. Anahory, "Tunable exchange bias in the magnetic weyl semimetal Co$_3$Sn$_2$S$_2$," Phys. Rev. B 105, 144423 (2022).

[158] J. Nogués and I. K. Schuller, "Exchange bias," J. Magn. Magn. Mater. 192, 203 (1999).

[159] S. Giri, M. Patra, and S. Majumdar, "Exchange bias effect in alloys and compounds," J. Phys. Condens. Matter 23, 073201 (2011).

[160] M. Ikhlas, T. Tomita, T. Koretsune, M.-T. Suzuki, D. Nishio-Hamane, R. Arita, Y. Otani, and S.

Nakatsuji, "Large anomalous Nernst effect at room temperature in a chiral antiferromagnet," Nat. Phys. 13, 1085 (2017).

[161] M. V. Berry, "Quantal phase factors accompanying adiabatic changes," Proc. R. Soc. Lond. A 392, 45 (1984).

[162] N. Nagaosa and Y. Tokura, "Topological properties and dynamics of magnetic skyrmions," Nat. Nanotechnol. 8, 899 (2013).

[163] S. Mühlbauer, B. Binz, F. Jonietz, C. Pfleiderer, A. Rosch, A. Neubauer, R. Georgii, and P. Bni, "Skyrmion lattice in a chiral magnet," Science 323, 915 (2009).

[164] P. Bruno, V. K. Dugaev, and M. Taillefumier, "Topological Hall effect and Berry phase in magnetic nanostructures," Phys. Rev. Lett. 93, 096806 (2004).

[165] A. Neubauer, C. Pfleiderer, B. Binz, A. Rosch, R. Ritz, P. G. Niklowitz, and P. Böni, "Topological Hall effect in the A phase of MnSi," Phys. Rev. Lett. 102, 186602 (2009).

[166] K. Ueda, S. Iguchi, T. Suzuki, S. Ishiwata, Y. Taguchi, and Y. Tokura, "Topological Hall effect in pyrochlore lattice with varying density of spin chirality," Phys. Rev. Lett. 108, 156601 (2012).

[167] S. E. Barnes and S. Maekawa, "Generalization of Faraday's law to include nonconservative spin forces," Phys. Rev. Lett. 98, 246601 (2007).

[168] N. Nagaosa, "Emergent inductor by spiral magnets," Jpn. J. Appl. Phys. 58, 120909 (2019).

[169] T. Yokouchi, F. Kagawa, M. Hirschberger, Y. Otani, N. Nagaosa, and Y. Tokura, "Emergent electromagnetic induction in a helical-spin magnet," Nature 586, 232 (2020).

[170] A. Kitaori, N. Kanazawa, T. Yokouchi, F. Kagawa, N. Nagaosa, and Y. Tokura, "Emergent electromagnetic induction beyond room temperature," P. Natl. Acad. Sci. 118, e2105422118 (2021).

[171] J. Ieda and Y. Yamane, "Intrinsic and extrinsic tunability of Rashba spin-orbit coupled emergent inductors," Phys. Rev. B 103, L100402 (2021).

[172] D. Kurebayashi and N. Nagaosa, "Electromagnetic response in spiral magnets and emergent inductance," Commun. Phys. 4, 260 (2021).

[173] M. M. Qazilbash, J. J. Hamlin, R. E. Baumbach, L. Zhang, D. J. Singh, M. B. Maple, and D. N. Basov, "Electronic correlations in the iron pnictides," Nat. Phys. 5, 647 (2009).

[174] Y. Shao, A. N. Rudenko, J. Hu, Z. Sun, Y. Zhu, S. Moon, A. J. Millis, S. Yuan, A. I. Lichtenstein, D. Smirnov, Z. Q. Mao, M. I. Katsnelson, and D. N. Basov, "Electronic correlations in nodal-line semimetals," Nat. Phys. 16, 636 (2020).

[175] A. Comanac, L. de' Medici, M. Capone, and A. J. Millis, "Optical conductivity and the correlation strength of high-temperature copper-oxide superconductors," Nat. Phys. 4, 287 (2008).

[176] G. Pokharel, S. M. L. Teicher, B. R. Ortiz, P. M. Sarte, G. Wu, S. Peng, J. He, R. Seshadri, and S. D. Wilson, "Electronic properties of the topological kagome metals $YV_6Sn_6$ and $GdV_6Sn_6$," Phys. Rev. B 104, 235139 (2021).

[177] H. Ishikawa, T. Yajima, M. Kawamura, H. Mitamura, and K. Kindo, "$GdV_6Sn_6$: A multi-carrier metal with non-magnetic 3d-electron kagome bands and 4f-electron magnetism," J. Phys. Soc. Jpn. 90, 124704 (2021).

... 
[178] G. Pokharel, B. Ortiz, P. Sarte, L. Kautzsch, G. Wu, J. Ruff, and S. D. Wilson, "Highly anisotropic magnetism in the vanadium-based kagome metal TbV$_6$Sn$_6$," arXiv preprint arXiv:2205.15559 (2022).

[179] J. Lee and E. Mun, "Anisotropic magnetic property of single crystals RV$_6$Sn$_6$ (R = Y, Gd-Tm, Lu)," arXiv preprint arXiv:2206.02924 (2022).

[180] X. Zhang, Z. Liu, Q. Cui, N. Wang, L. Shi, H. Zhang, X. Dong, J. Sun, Z. Dun, and J. Cheng, "Electronic and magnetic properties of intermetallic kagome magnets RV$_6$Sn$_6$ (R = Tb-Tm)," arXiv preprint arXiv:2206.05653 (2022).

[181] S. V. Isakov, S. Wessel, R. G. Melko, K. Sengupta, and Y. B. Kim, "Hard-core bosons on the kagome lattice: Valence-bond solids and their quantum melting," Phys. Rev. Lett. 97, 147202 (2006).

[182] S. Nishimoto, M. Nakamura, A. O'Brien, and P. Fulde, "Metal-insulator transition of fermions on a kagome lattice at 1/3 filling," Phys. Rev. Lett. 104, 196401 (2010).

[183] A. Rüegg and G. A. Fiete, "Fractionally charged topological point defects on the kagome lattice," Phys. Rev. B 83, 165118 (2011).

[184] W.-S. Wang, Z.-Z. Li, Y.-Y. Xiang, and Q.-H. Wang, "Competing electronic orders on kagome lattices at van Hove filling," Phys. Rev. B 87, 115135 (2013).

[185] Y. Wang, G. T. McCandless, X. Wang, K. Thanabalasingam, H. Wu, D. Bouwmeester, H. S. J. van der Zant, M. N. Ali, and J. Y. Chan, "Electronic properties and phase transition in the kagome metal Yb$_{0.5}$Co$_3$Ge$_3$," Chem. Mater. 34, 7337 (2022).

## Acknowledgments

This work was supported by the National Key R&D Program of China grant number 2022YFA1403603, National Natural Science Foundation of China grant numbers 12225401, U1832214, U2032213 and 12104461, and Strategic Priority Research Program of Chinese Academy of Sciences grant number XDB28000000. A portion of this work was supported by the High Magnetic Field Laboratory of Anhui Province. X.X. acknowledges support from the China Postdoctoral Science Foundation grant number 2022T150655 and 2020M682056, Anhui Postdoctoral Foundation grant number 2020B472, Anhui Provincial Natural Science Foundation grant number 2108085QA23, the HFIPS Director's Fund grant number YZJJ2021QN28, and Special Research Assistant Program, Chinese Academy of Sciences.

## Competing interests

The authors declare no competing interests.